\documentclass[
reprint,
superscriptaddress,
amsmath,
amssymb,
aps,
prl,
floatfix,
]{revtex4-2}
\usepackage{graphicx}
\usepackage{amsmath}
\usepackage{hyperref}
\usepackage{xcolor}
\usepackage{multirow}

\begin{document}

\title{Loss of altermagnetic order and smooth restoration of Kramers'\\spin degeneracy with increasing temperature in CrSb and MnTe}

\author{Christopher D. Woodgate}
\thanks{Present address: Yusuf Hamied Department of Chemistry, University of Cambridge, Lensfield Road, Cambridge, CB2 1EW, United Kingdom}
\email{cw2016@cam.ac.uk}
\affiliation{H.H. Wills Physics Laboratory, University of Bristol, Royal Fort, Bristol, BS8 1TL, United Kingdom}
\affiliation{Department of Physics, University of Warwick, Coventry, CV4 7AL, United Kingdom}
\author{Nabil Menai}
\affiliation{H.H. Wills Physics Laboratory, University of Bristol, Royal Fort, Bristol, BS8 1TL, United Kingdom}
\author{Arthur Ernst}
\affiliation{Institute for Theoretical Physics, Johannes Kepler University Linz, Altenberger Strasse 69, A-4040 Linz, Austria}
\affiliation{Donostia International Physics Center, Donostia-San Sebastian 20018 Gipuzkoa, Spain}
\affiliation{Max-Planck-Institut f\"ur Mikrostrukturphysik, Weinberg 2, D-06120 Halle, Germany}
\author{Julie B. Staunton}
\email{J.B.Staunton@warwick.ac.uk}
\affiliation{Department of Physics, University of Warwick, Coventry, CV4 7AL, United Kingdom}

\begin{abstract}
We describe how thermally induced spin fluctuations modify the electronic structures of two prototypical altermagnets, CrSb and MnTe, via application of the disordered local moment picture.
For both materials, our self-consistent, \textit{ab initio} calculations demonstrate that local magnetic moments persist on Cr and Mn atoms in their paramagnetic states, necessitating a spin-polarised description of the electronic structure even above the N\'eel temperature, $T_\mathrm{N}$.
Moreover, Kramers' spin degeneracy, which is broken for both materials in their altermagnetic ground states, is shown to be smoothly restored -- on the average -- as the local moments thermally disorder.
In metallic CrSb, this occurs at temperatures well below $T_\mathrm{N}$ and the signature effects of its altermagnetism are lost as the magnetic disorder induces heavy smearing of strongly dispersive electronic states around the Fermi energy.  
By contrast, in semiconducting MnTe, with its band gap largely unaffected by magnetic disorder, the spin degeneracy only returns at temperatures close to and above $T_\mathrm{N}$. 
We quantify the temperature dependence of the altermagnetic order parameter and the underlying electronic structures of both materials, with implications for their spin transport {and spectroscopic} properties.
\end{abstract}

\date{November 3, 2026}

\maketitle

Altermagnetism has emerged as a new class of collinear magnetic order -- distinct from all three of ferromagnetism, antiferromagnetism, and ferrimagnetism. Altermagnets are materials in which there is a net-zero spin polarisation, akin to antiferromagnets, but in which there is spin splitting of electronic bands across large portions of the Brillouin zone, similar to ferromagnets and ferrimagnets~\cite{smejkal_emerging_2022}. In the absence of spin-orbit coupling, this broken Kramers' spin degeneracy has its origins in the underlying symmetries of the crystal structure, particularly as enforced by the presence of non-magnetic atoms~\cite{smejkal_beyond_2022, ahn_antiferromagnetism_2019}. A growing number of materials have now been experimentally verified to be altermagnets. Experimental signatures of the phenomenon include those found spectroscopically~\cite{krempasky_altermagnetic_2024, reimers_direct_2024, osumi_observation_2024, ding_large_2024, hajlaoui_temperature_2024, amin_nanoscale_2024, yang_three-dimensional_2025, santhosh_altermagnetic_2025, galindezruales_revealing_2025, pan_experimental_2026}, those found in electronic transport measurements~\cite{feng_anomalous_2022, bai_observation_2022, karube_observation_2022, bai_efficient_2023, gonzalez_betancourt_spontaneous_2023, reichlova_observation_2024, galindezruales_revealing_2025}, and also those found in magnetic excitation spectra~\cite{liu_chiral_2024, sun_observation_2025, singh_coherent_nodate}.

{Altermagnetic materials are of interest both from a fundamental physical perspective, as well as for their potential application as functional materials in fields such as spintronics~\cite{bai_altermagnetism_2024, song_altermagnets_2025}. A key aspect to follow from both of these perspectives is how altermagnetic order diminishes with increasing temperature, $T$. In this Letter, we describe the effects of thermally induced spin fluctuations on the spin-polarised electronic structure of altermagnets. As exemplars, we study two prototypical $g$-wave altermagnets: CrSb, which is metallic, and MnTe, which is semiconducting. Starting with descriptions of two limiting cases: (i) perfect altermagnetic order at $T=0$~K and (ii) complete magnetic disorder above the N\'eel temperature for $T>T_\mathrm{N}$, we then track how the Kramers' spin degeneracy is restored as the long-range altermagnetic order diminishes with rising temperature. Notably, we find signatures of changes to the transport lifetimes and mean free paths caused by the thermally induced spin disorder. Our results, therefore, have significant implications for the temperature dependence of the materials' spin transport and spectroscopic properties.

Within the framework of theoretical calculations, to demonstrate how altermagnetic ordering results in non-relativistic spin splitting of bands in a material, comparisons are often made between the band structures  of a non-magnetic state and the altermagnetic ground state. However, in many magnetic materials, the description of the paramagnetic state as one which is completely non-magnetic is unphysical~\cite{staunton_electronic_1985}, yielding estimates of magnetic transition temperatures which are much too high~\cite{gunnarsson_band_1976}. Instead, appropriate `spins', or `local moments', should be identified that fluctuate thermally. Collective electron spin effects emerge on essentially two different time scales.  On a fast time scale, electrons cooperate to form atomic-scale regions of spin polarised density and, on a longer time scale, these local moments slowly vary their orientations~\cite{gyorffy_first-principles_1985}. 
The disordered local moment (DLM) picture of magnetism at finite temperature~\cite{pindor_disordered_1983, staunton_disordered_1984, gyorffy_first-principles_1985, hughes_onset_2008} is based on this insight and considers the statistical physics of the local moments.}

\begin{figure*}[t]
    \centering
    \includegraphics[width=\textwidth]{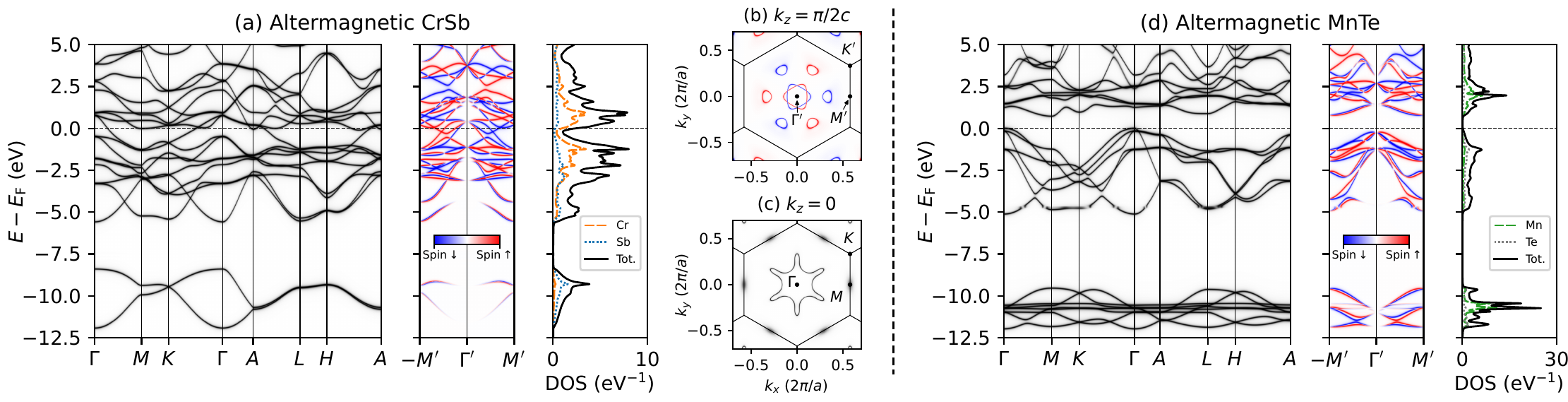}
    \caption{Comparison of the electronic structure of CrSb (a--c) and MnTe (d) in their altermagnetic ground states at $T=0$~K. We show both the band structure and total electronic density of states. For metallic CrSb, we also show two slices of the Fermi surface (b,c). In both materials, altermagnetic band splitting is visible along the $-M'$-$\Gamma'$-$M'$ path in reciprocal space.}
    \label{fig:altermagnetic_band_structure}
\end{figure*}

{For the paramagnetic state of the archetypal metallic ferromagnet, bcc Fe, the DLM picture produces an electronic structure where there is local spin splitting which is wavevector- and energy-dependent~\cite{staunton_electronic_1985}. This is in agreement with seminal spectroscopic experiments~\cite{kirschner_wave-vector-dependent_1984, kisker_temperature_1984}, where it is found that, in some regions of the Brillouin zone, a local spin polarisation is well-defined and distinct, whereas in other places it is not evident at all or there is strong disorder broadening of the states. Here we find analogous temperature-dependent features in the paramagnetic state of altermagnetic CrSb and MnTe, in which sizeable local moments also persist above $T_\mathrm{N}$. While the electronic structure of such a paramagnetic state, when thermally averaged over all the local moment configurations, has zero overall spin polarisation, it differs qualitatively from the electronic structure of a non-magnetic state owing to the spin disorder~\cite{staunton_electronic_1985}. For intervening, partially altermagnetically ordered states, $ 0 < T < T_\mathrm{N}$, we show that changes to the spin-polarised electronic structure encapsulate the effects of the spin disorder and lead to the signatures of broken Kramers' spin degeneracy being lost before $T_\mathrm{N}$ is reached.}

We begin by modelling the electronic structure of both materials in their altermagnetic $T=0$~K ground states. We use the all-electron \textsc{Hutsepot} code~\cite{hoffmann_magnetic_2020} to construct the self-consistent potentials of the Korringa--Kohn--Rostoker (KKR) formulation~\cite{ebert_calculating_2011, faulkner_multiple_2018} of density functional theory (DFT)~\cite{jones_density_2015}. For both materials, we use experimentally determined lattice constants and perform scalar relativistic calculations to focus on the defining feature of `strong' altermagnetism: exchange-splitting of electronic bands in the absence of spin-orbit coupling~\cite{cheong_altermagnetism_2025}. For CrSb, which is metallic, we use the local spin-density approximation (LSDA) to the DFT exchange-correlation functional~\cite{perdew_accurate_1992}. For MnTe, which is semiconducting, we employ the local self-interaction correction (LSIC)~\cite{luders_self-interaction_2005}  to the LSDA to capture {this material's strong electronic correlations in a parameter-free way}. Full details of all calculations are given in the Supplemental Material~\cite{supplemental}.

\begin{figure*}[t]
    \centering
    \includegraphics[width=\textwidth]{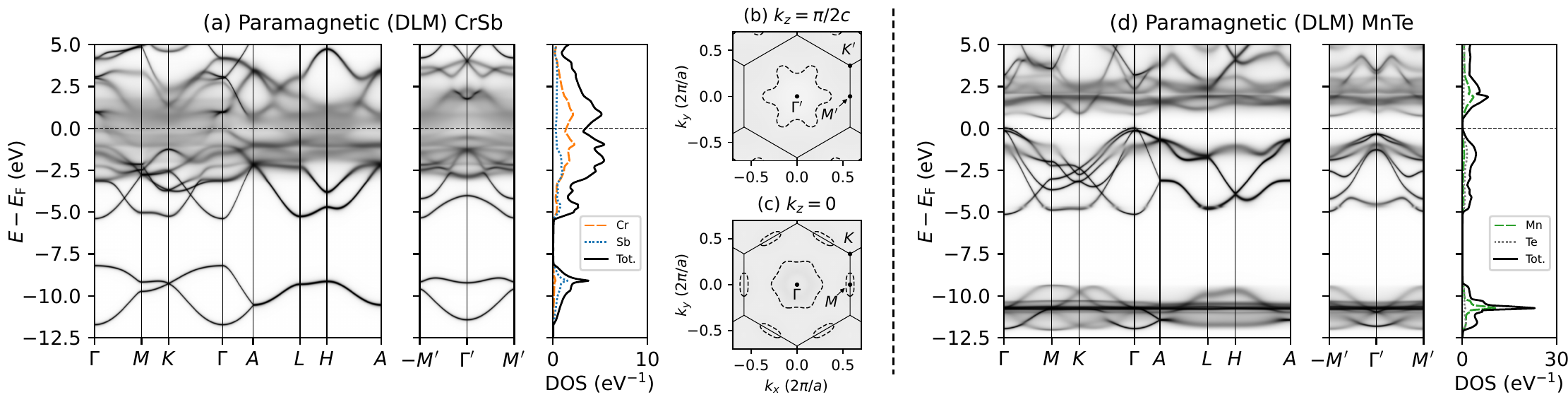}
    \caption{Comparison of the electronic structure of CrSb (a--c) and MnTe (d) in their paramagnetic (DLM) states. We show both the BSF and total electronic density of states, both of which are heavily smeared at certain points by the magnetic disorder. For metallic CrSb, we also show two slices of the Fermi surface with dashed contours of constant spectral weight. Because the Fermi surface is washed out by the magnetic disorder, we plot a contour of constant spectral weight, evidencing the six-fold degeneracy in both planes. In MnTe, which is semiconducting, the band gap is largely unaffected by the magnetic disorder. Where band splitting was evidenced in Fig.~\ref{fig:altermagnetic_band_structure}, we now show the total BSF in greyscale, as it is now fully spin-degenerate.}
    \label{fig:paramagnetic_band_structure}
\end{figure*}

Fig.~\ref{fig:altermagnetic_band_structure} shows the Bloch spectral function (BSF), $A_\mathrm{B}(E, \mathbf{k})$, and electronic density of states (DOS) for both CrSb and MnTe in their altermagnetic ground states. The BSF, which reduces to the conventional band structure in a pristine material at $T=0$~K~\cite{ebert_calculating_2011}, is shown along the conventional high-symmetry path, as well as along the $-M'$-$\Gamma'$-$M'$ path to evidence the altermagnetic band splitting~\footnote{Strictly, along these low-symmetry paths, we plot the difference between spin-up and spin-down BSFs, \textit{i.e.} the quantity $A^\uparrow_\mathrm{B}(E, \mathbf{k}) - A^\downarrow_\mathrm{B}(E, \mathbf{k})$.}. For CrSb, which is metallic in our calculations, we also plot two slices of the Fermi surface in the $k_z = 0$ and $k_z= \pi/2c$ planes. In both materials, along the low-symmetry path, there is clear spin splitting of the electronic band structure. In CrSb, this spin splitting can also be seen in the material's Fermi surface, with the slice in the $k_z = 0$ plane being spin-degenerate and exhibiting six-fold rotational symmetry, while the slice in the $k_z= \pi/2c$ plane shows spin splitting and three-fold rotational symmetry (with six-fold rotational symmetry relating spin-up and spin-down electrons). In MnTe, which is semiconducting, the LSIC-LSDA recovers a band gap -- the indirect gap is 0.6~eV, while the direct gap at the $\Gamma$ point is 1.2~eV. These values are in fair agreement with literature~\cite{allen_optical_1977, podgorny_electronic_1983, ferrer-roca_temperature_2000}. The magnitude of the local magnetic moments on Cr atoms in CrSb is 2.70~$\mu_\mathrm{B}$, while for Mn atoms in MnTe this value is 4.62~$\mu_\mathrm{B}$, {in agreement with experimental determinations of $2.7\pm 0.2$~$\mu_\mathrm{B}$ for CrSb~\cite{snow_neutron_1952} and 4.6~$\mu_\mathrm{B}$ for MnTe~\cite{kunitomi_neutron_1964}}.

Proceeding, we now consider the influence of magnetic disorder, and present the results of calculations for both materials in the paramagnetic regime, \textit{i.e.} above $T_\mathrm{N}$. Practically, the average over all possible orientations of magnetic moment for each lattice site is achieved via application of the coherent potential approximation (CPA)~\cite{soven_coherent-potential_1967, stocks_complete_1978}, which facilitates recovery of the average Green's function for a disordered system within a single-site approximation. With the same DFT settings as for the altermagnetic calculations outlined above, we perform self-consistent calculations on CrSb and MnTe in their paramagnetic (DLM) states. For both materials, we find that the magnetic atoms support substantial local magnetic moments in this paramagnetic regime, with the magnitude of the local magnetic moment on Cr atoms in CrSb calculated to be 2.58 $\mu_\mathrm{B}$, and for Mn atoms in MnTe calculated to be 4.63 $\mu_\mathrm{B}$. That the magnitudes of magnetic moments are largely unchanged between altermagnetic and paramagnetic states confirms that these are `good' local moments~\cite{staunton_using_2014, patrick_marmot_2022}. We calculate $T_\mathrm{N}$ to be 920~K and 160~K for CrSb and MnTe respectively, in fair agreement with the values 705~K~\cite{takei_magnetic_1963} and 306~K~\cite{reig_growth_2001} determined experimentally. For MnTe, we attribute the underestimation of $T_\mathrm{N}$ to the strong localisation of $3d$ states within the LSIC-LSDA. If we use the LSDA alone, MnTe is found to be metallic, and $T_\mathrm{N}$ is instead overestimated to be 709~K.

\begin{figure*}[t]
    \centering
    \includegraphics[width=\textwidth]{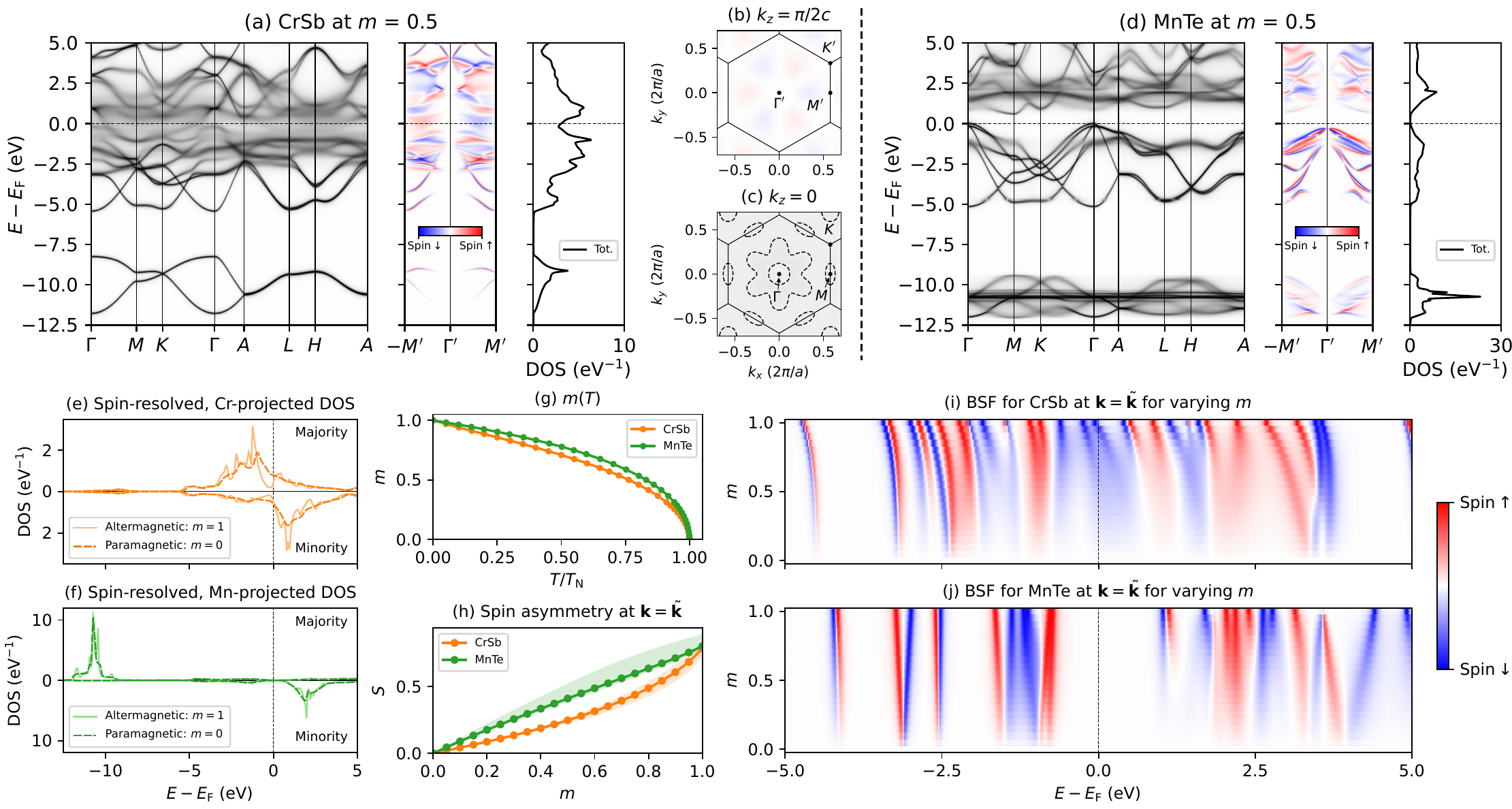}
    \caption{Evolution of the electronic structures of CrSb and MnTe with temperature, $T$, and magnetic order, $m$. Panels (a--c) and (d) show the BSF, DOS and Fermi surface (for CrSb) of CrSb and MnTe, respectively, at $m=0.5$. This value of $m$ corresponds to $T/T_\mathrm{N}\approx0.76$ and 0.83 for the two materials, respectively. Panels (e) and (f) show the spin-resolved DOS of Cr and Mn in the limiting cases of $m=0$ and $m=1$. Panel (g) then shows $m(T)$ for both materials, while panel (h) plots the spin asymmetry parameter, $S(m)$. Finally, panels (i) and (j) show a spin-resolved, constant-$\mathbf{k}$ slice of the BSF for both materials as a function of $m$ around $E_\mathrm{F}$ at $\tilde{\mathbf{k}} = (\frac{2\pi\sqrt{3}}{3a}, 0, \frac{\pi}{2c})$, halfway between the points $\Gamma'$ and $M'$. Though local moments persist in the paramagnetic state, altermagnetic spin splitting of bands rapidly diminishes with decreasing $m$.}
    \label{fig:transition}
\end{figure*}

Plots of the BSF and electronic DOS for both materials, as well as Fermi surface cuts for CrSb, simulated in this paramagnetic state are shown in Fig.~\ref{fig:paramagnetic_band_structure}. {As a consequence of the spin disorder, the energy dependence of the BSF at a particular wavevector now resembles a set of Lorentzian-type peaks where the reciprocals of breadths of those peaks capture the lifetimes of the electronic states. In a similar vein, for a fixed energy, for example the Fermi energy in CrSb, the loci of the peaks of the BSF map out a surface in wavevector space where now the reciprocal of the breadth of a peak provides information directly about the electronic mean free path. A guideline for assessing what magnitude of broadening to the BSF to expect at some wavevector and energy is based on the ratio between the spin splitting and the band dispersion. Large splittings and narrow bands favour modest broadening whereas small values and wide bands lead to more smeared structures~\cite{staunton_electronic_1985}.}

{For both materials, in this paramagnetic state in which all orientations of magnetic moment are equally probable, the spin splitting of the band structure along the low-symmetry path seen in Fig.~\ref{fig:altermagnetic_band_structure}  \textit{completely disappears}. Moreover, a six-fold symmetry in the $k_z= \pi/2c$ cut of the Fermi surface of CrSb is restored. For MnTe, the magnetic disorder does not substantially affect the band gap predicted in this model, with the indirect gap now being 0.6 eV.} That the band gap of MnTe is largely unaffected by the magnetic disorder is consistent with previous calculations simulating the paramagnetic states of $3d$ transition metal oxides, including MnO~\cite{hughes_onset_2008}. Though the collapse of relativistic spin splitting -- so-called `weak' altermagnetism -- due to magnetic disorder has been described previously~\cite{hajlaoui_temperature_2024}, here we extend this result to the case of non-relativistic spin splitting, \textit{i.e.} `strong' altermagnetism. Additionally, the magnetic disorder induces changes and heavy smearing of the band structure at certain energies and in portions of the Brillouin zone. Indeed, some fine details in the band structures of both materials are no longer resolvable. A band structure symmetry analysis is given in the Supplemental Material~\cite{supplemental}.

We now focus on our key result: how these altermagnets' respective electronic structures evolve as they are heated up towards their N\'eel temperatures, and how Kramers' spin degeneracy is smoothly restored as the local moments thermally disorder. We introduce a magnetic order parameter, $m$, to interpolate between the two limiting cases of altermagnetism ($m=1$) and paramagnetism ($m=0$). {Panels (a--c) and (d) of Fig.~\ref{fig:transition} show the electronic structures of CrSb and MnTe at $m=0.5$, reflective of a situation in which there is partial altermagnetic order, as might be observed experimentally. The smearing of the bands and, in particular, the washing out of the spin-split features along the low-symmetry paths and Fermi surface cuts emphasise that long-ranged magnetic order need not vanish entirely for significant impact to be felt. Panels (e) and (f)} compare the spin- and species-resolved DOS projected onto a single Cr or Mn atom in the limits $m=1$ and $m=0$. These plots are spin-resolved, with the local orientation of a given magnetic moment defining the axis of quantisation. Associated with the local magnetic moments obtained in these self-consistent calculations, there is substantial local exchange-splitting of electronic states, which persists in the paramagnetic state for both materials. {Panel (g)} then displays the temperature evolution of $m$ as a function of $T$ as predicted within the DLM picture using the \textsc{Marmot} package~\cite{staunton_temperature_2006, patrick_marmot_2022, supplemental}. {The overall observed $m(T)$ trend for both materials is broadly consistent with experimental determinations~\cite{singh_coherent_nodate, kriegner_magnetic_2017}, though it should be stressed that the results presented here are mean-field and therefore perfect agreement is not anticipated. Continuing, panels (i) and (j) evidence} how the electronic structure of the selected materials evolves during this transition via a cut of the spin-resolved BSF around the Fermi energy, $E_\mathrm{F}$, at the point $\tilde{\mathbf{k}} = (\frac{2\pi\sqrt{3}}{3a}, 0, \frac{\pi}{2c})$. This point is halfway between $\Gamma'$ and $M'$, where we find that there is substantial spin splitting of bands for both materials in the $T=0$~K altermagnetic ground state. {Panel (h) serves to quantify panels (i) and (j) by introducing the \textit{spin asymmetry} at $\tilde{\mathbf{k}}$ for varying $m$, 
\begin{equation}
S(m)=\frac{\int \mathrm{d}E\,w(E)\,\bigl|A^{\uparrow}_{\mathrm{B}}(\tilde{\mathbf{k}},E;m)-A^{\downarrow}_{\mathrm{B}}(\tilde{\mathbf{k}},E;m)\bigr|}
          {\int \mathrm{d}E\,w(E)\,\bigl[A^{\uparrow}_{\mathrm{B}}(\tilde{\mathbf{k}},E;m)+A^{\downarrow}_{\mathrm{B}}(\tilde{\mathbf{k}},E;m)\bigr]},
\end{equation}
where $A^{\sigma}_{\mathrm{B}}$ is the spin-resolved Bloch spectral function and $w(E)=\exp[-(E-E_{\mathrm{F}})^{2}/2\sigma^{2}]$, with $\sigma=0.75$~eV, selects the states near $E_{\mathrm{F}}$ that govern transport. $S$ vanishes identically at $m=0$ by symmetry, and is independent of the normalisation of $w$. The shaded bands in Fig.~\ref{fig:transition}(h) indicate the sensitivity of $S$ to the window function choice, discussed in the Supplemental Material~\cite{supplemental}.}

Here, several key observations should be made. First is that, for both materials, even a small degree of magnetic disorder induces significant smearing of electronic bands at some energies, as well as changes to the band structure, group velocity, and a reduction in the degree of band splitting. {Such findings can be compared with temperature-dependent ARPES and transport data, \textit{e.g.} Refs.~\cite{hajlaoui_temperature_2024, naduvile_thadathil_multicarrier_2026}.} Second is that this process progresses more rapidly in metallic CrSb, where {moderately dispersive} $3d$ states are found at energies around $E_\mathrm{F}$, as compared to semiconducting MnTe, where {flatter bands of} $3d$ states are found split away to energies far above/below $E_\mathrm{F}$ in our LSIC-LSDA calculations. {That the presence of $3d$ states near the Fermi level results in more rapid loss of spin splitting is further evidenced in the Supplemental Material, where we show results for MnTe within both the LSDA and Slater transition state approximation; only when $3d$ states are split away from $E_\mathrm{F}$ does this robustness to magnetic disorder hold. In CrSb, the location of the spin-split bands relative to the Fermi level is altered by the magnetic disorder, while in MnTe they remain comparatively static.} Moreover, our calculations suggest that the altermagnetic order parameter, $m$, tails off more rapidly initially with increasing temperature (as a fraction of $T_\mathrm{N}$) for CrSb than it does for MnTe. {The smearing of electronic bands is directly related to reduced electronic mean free path and lifetimes in a material}, and favourable spin transport properties are conferred when band splitting in a material is large. It is evident, therefore, that thermally induced spin fluctuations {will have a significant} effect on the temperature dependence of these altermagnets' attractive electronic transport properties for spintronic applications.

In summary, we have considered the application of the disordered local moment picture to the study of two prototypical altermagnets, CrSb and MnTe, describing how thermally induced spin fluctuations modify their electronic structures. We predict that, for both materials above $T_\mathrm{N}$, localised magnetic moments persist on magnetic atoms, but that the spin-degeneracy of the materials' respective band structures is restored -- on the average -- by thermally induced spin fluctuations and associated site-diagonal spin disorder. This magnetic disorder also induces marked changes to, and heavy smearing of, the band structure of both materials, with the Fermi surface of CrSb experiencing significant qualitative changes and disorder broadening. However, in MnTe, the band gap of the material remains largely unaffected. We have then used our \textit{ab initio} modelling framework to predict $T_\mathrm{N}$ for both materials, as well as the temperature dependence of both materials' altermagnetic order parameter, $m$. Most significantly, we have described the evolution of both materials' electronic structures with increasing temperature. We assert that the inclusion of the thermally induced spin fluctuations which we describe will be essential when seeking to accurately model the spin transport properties of these materials, as well as for robust interpretation of experimental spectroscopy data. Our results also lay the foundations for future, fully relativistic studies (see, \textit{e.g.} Ref.~\citenum{ebert_calculating_2015}) examining the effect of spin fluctuations on the electronic, magnetic, and transport properties of these intriguing materials.

\begin{acknowledgments}
C.D.W.\ acknowledges support from EPSRC Grant EP/W524414/1. J.B.S.\ acknowledges support from EPSRC Grant EP/W021331/1. Computing resources were provided by the Advanced Computing Research Centre of the University of Bristol, and by the Scientific Computing Research Technology Platform of the University of Warwick. {The authors also acknowledge the use of compute resources provided by the Isambard 3 high-performance computing (HPC) facility. Isambard 3 is hosted by the University of Bristol and operated by the GW4 Alliance (\href{https://gw4.ac.uk}{https://gw4.ac.uk}) and is funded by UK Research and Innovation; and the EPSRC, Grant EP/X039137/1.}
\end{acknowledgments}

The data that support the findings of this article are publicly available~\cite{dataset}.

\end{document}


\title{Loss of altermagnetic order and smooth restoration of Kramers'\\spin degeneracy with increasing temperature in CrSb and MnTe\\\vspace{4pt}Supplemental Material}

\author{Christopher D. Woodgate}
\thanks{Present address: Yusuf Hamied Department of Chemistry, University of Cambridge, Lensfield Road, Cambridge, CB2 1EW, United Kingdom}
\email{cw2016@cam.ac.uk}
\affiliation{H.H. Wills Physics Laboratory, University of Bristol, Royal Fort, Bristol, BS8 1TL, United Kingdom}
\affiliation{Department of Physics, University of Warwick, Coventry, CV4 7AL, United Kingdom}
\author{Nabil Menai}
\affiliation{H.H. Wills Physics Laboratory, University of Bristol, Royal Fort, Bristol, BS8 1TL, United Kingdom}
\author{Arthur Ernst}
\affiliation{Institute for Theoretical Physics, Johannes Kepler University Linz, Altenberger Strasse 69, A-4040 Linz, Austria}
\affiliation{Donostia International Physics Center, Donostia-San Sebastian 20018 Gipuzkoa, Spain}
\affiliation{Max-Planck-Institut f\"ur Mikrostrukturphysik, Weinberg 2, D-06120 Halle, Germany}
\author{Julie B. Staunton}
\email{J.B.Staunton@warwick.ac.uk}
\affiliation{Department of Physics, University of Warwick, Coventry, CV4 7AL, United Kingdom}

\maketitle

This is the Supplemental Material accompanying the main text. Here, we present additional discussion, results, and analyses to aid interpretation of our data, and also report the technical details of our calculations.

\section{The disordered local moment picture of finite-temperature magnetism}
\label{suppsec:dlm}
The theoretical framework we use has its foundations in the seminal paper by Gy\H{o}rffy \textit{et al.}~\cite{gyorffy_first-principles_1985} on ferromagnetic phase transitions in transition metals, and has subsequently been generalised to strongly correlated materials such as transition metal oxides~\cite{hughes_onset_2008}, as well as to systems with multiple magnetic atoms in the unit cell~\cite{mendive-tapia_ab_2019}.  The disordered local moment (DLM) picture of magnetism at finite temperature~\cite{pindor_disordered_1983, staunton_disordered_1984, gyorffy_first-principles_1985, hughes_onset_2008} considers the statistical physics of the thermally fluctuating, atomically localised magnetic moments which characterise the relatively slowly evolving collective electron spin degrees of freedom in a magnetic material. The picture is equally applicable to metallic, semiconducting and insulating systems. Fundamentally, the only requirement a system must fulfil for application of the DLM picture to be appropriate is that atomically localised magnetic moments persist on magnetic atoms above the magnetic critical temperature. In this work, both metallic CrSb and semiconducting MnTe fulfil this requirement. 

The DLM approach begins with consideration of the grand potential, $\Omega$, of a system of interacting electrons where the local moment configuration is constrained, as can be described within the tenets of density functional theory (DFT). An \textit{ab initio} expression for the internal energy of a magnetic material~\cite{gyorffy_first-principles_1985, hughes_onset_2008, mendive-tapia_ab_2019} is then given by the average of this quantity, $\bar{\Omega}$, over an ensemble of local moment orientational configurations.
The orientation of such a local moment at site $i$ with combined lattice ($I$) and basis ($s$) index $i=(I,s)$ is described by a unit vector, $\hat{\mathbf{e}}_{i}$, defined via
\begin{equation}
    \hat{\mathbf{e}}_{i} = \frac{\boldsymbol{\mu}_i}{\|\boldsymbol{\mu}_i\|}, \qquad \boldsymbol{\mu}_i = \int_{V_i} \textrm{d} \mathbf{r}\;\mathbf{M}(\mathbf{r}),
    \label{eq:local_moment_definition}
\end{equation}
where $\boldsymbol{\mu}_i$ is a vector specifying the size and orientation of a magnetic moment at a particular site, obtained by integrating the usual electron spin density $\mathbf{M}(\mathbf{r})$ over the surrounding Wigner-Seitz volume, $V_i$. In the magnetic ground state of a magnetic material, in the zero-temperature limit, the set $\{\hat{\mathbf{e}}_i\}$ defines the relative arrangement of these local moments with respect to one another (\textit{i.e.} the magnetic order), and their orientation with respect to the underlying crystal lattice. However, at finite temperature, thermally induced spin fluctuations cause the $\{\hat{\mathbf{e}}_i\}$ to deviate from this ground-state arrangement, and the likelihood of any arrangement of local moment orientations is determined by a probability distribution, $P(\{\hat{\mathbf{e}}_i\})$. In general, the magnetic order parameters, $\{\mathbf{m}_i\}$, are averages of the $\{\hat{\mathbf{e}}_i\}$ using this probability distribution. At $T=0$~K these order parameters have magnitude one and in the paramagnetic state zero. For CrSb and MnTe, where the magnetic ground state is collinear and the two magnetic sublattices have equal and opposite magnetisations, the magnitude of the magnetic order parameter on one site functions as a global altermagnetic order parameter for the materials. The probability distribution, $P(\{\hat{\mathbf{e}}_i\})$, is given in terms of {\it ab initio} Weiss fields dependent on $\bar{\Omega}$, the local moment configurational average of the grand potential~\cite{gyorffy_first-principles_1985, hughes_onset_2008, mendive-tapia_ab_2019}. 

Above a material's magnetic critical temperature, in the high-temperature limit, the Weiss fields vanish and magnetic moments are decorrelated from one another and have no preferred direction. In this case, the probability distribution for a given local moment $\hat{\mathbf{e}}_i$ is simply
\begin{equation}
    P(\hat{\mathbf{e}}_i) = \frac{1}{4\pi}, \label{eq:paramagnetic_probability}
\end{equation}
\textit{i.e.} it is uniform on the sphere. Here, while it may be the case that $\|\boldsymbol{\mu}_i\|$ remains finite, the net magnetisation of the material is zero once the average is taken over the ensemble of all possible local moment orientations. Determination of the magnetic critical temperature by considering the static, paramagnetic spin susceptibility is achieved following the prescription of Ref.~\cite{staunton_static_1986} and discussed further below, in Sec.~\ref{suppsec:spin_susceptibility}. Determination of the smooth evolution of this probability distribution (approaching $T_\mathrm{N}$ from below) and, thus, the altermagnetic order parameter and electronic structure, is achieved following the prescription of Refs.~\citenum{staunton_temperature_2004} and \citenum{staunton_temperature_2006} and discussed further below, in Sec.~\ref{suppsec:magnetic_order}.

\section{The N\lowercase{i}A\lowercase{s} crystal structure and its Brillouin zone}

\begin{figure}[h]
    \centering
    \includegraphics[width=0.6\linewidth]{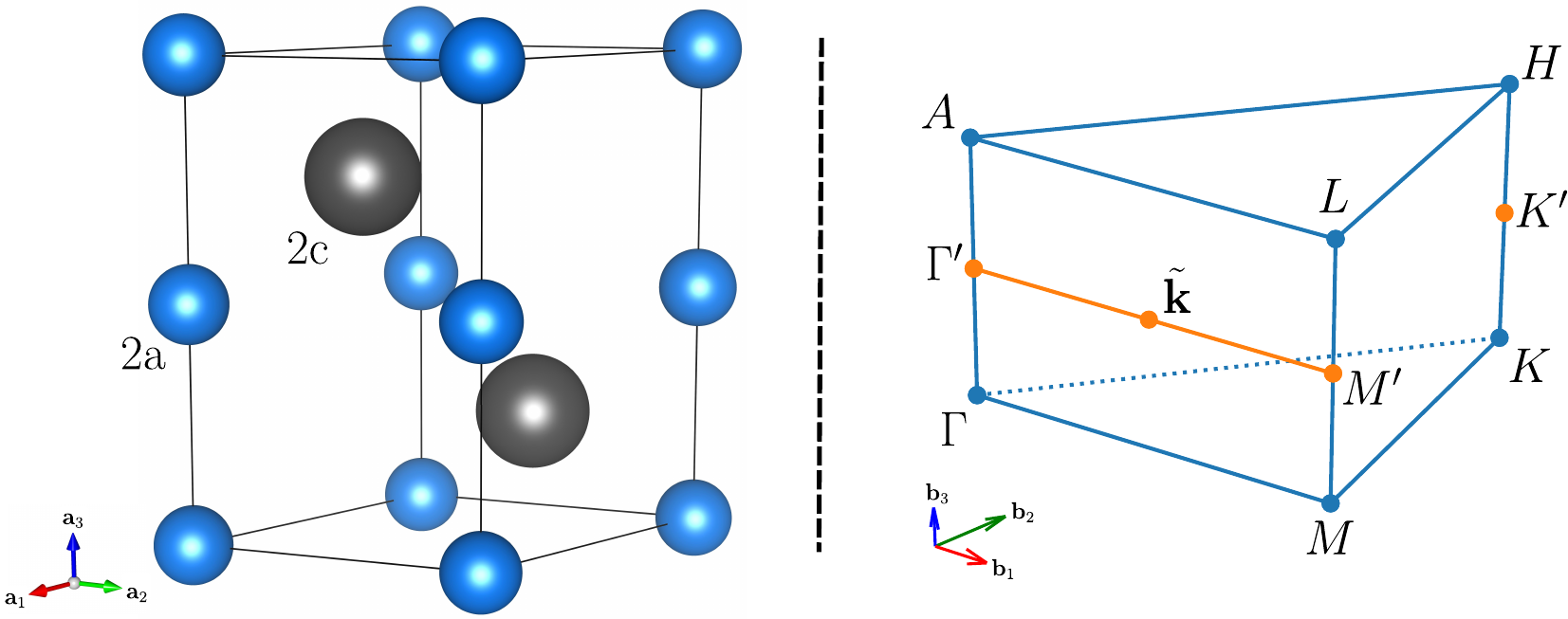}
    \caption{Illustration of the real-space crystal structure (left) and irreducible wedge of the Brillouin zone (right) of the hexagonal NiAs structure, space group $P6_3/mmc$ (No.~194) in which both CrSb and MnTe crystallise. We label the special points of the Brillouin zone, including selected lower-symmetry points where the altermagnetic band splitting is evidenced. The point $\tilde{\mathbf{k}}$ is chosen to lie half way between $\Gamma'$ and $M'$. The image on the left is generated using \textsc{VESTA}~\cite{momma_vesta_2011}.}
    \label{fig:nias_structure_ibz}
\end{figure}

In Fig.~\ref{fig:nias_structure_ibz}, we illustrate the NiAs crystal structure, space group $P6_3/mmc$ (No.~194), and its associated irreducible Brillouin zone. (The NiAs structure is the structure in which CrSb and MnTe both crystallise.) We explicitly show the low-symmetry $\Gamma'$-$M'$ path, used to evidence the altermagnetic band splitting (or lack thereof) in our calculations. We also show the location of $K'$. $\Gamma'$ is positioned halfway between $\Gamma$ and $A$, $M'$ is positioned halfway between $M$ and $L$, and $K'$ is positioned halfway between $K$ and $H$. For Figure 3 of the main text, and for Figures \ref{fig:mnte_lsda_evolution} and \ref{fig:mnte_sts_evolution} in this document, a constant-$\mathbf{k}$ slice of the BSF for varying altermagnetic order parameter, $m$, is shown at $\mathbf{k} = (\frac{2\pi\sqrt{3}}{3a}, 0, \frac{\pi}{2c})$. This point lies half way between $\Gamma'$ and $M'$, and we label it $\tilde{\mathbf{k}}$ for notational convenience.

Explicitly, the real-space lattice vectors of the NiAs structure (with hexagonal lattice parameters $a$ and $c$) are
\begin{equation}
        \mathbf{a}_1 = a\cdot(\frac{\sqrt{3}}{2},\quad -\frac{1}{2},\quad 0),\qquad
        \mathbf{a}_2 = a\cdot(0,\quad 1,\quad 0),\quad \textrm{and} \qquad
        \mathbf{a}_3 = c\cdot(0,\quad 0,\quad 1),
\end{equation}
while the basis vectors (specifying fractional atomic coordinates) and their respective Wyckoff labels are
\begin{equation}
        \mathbf{B}_1 = \mathbf{0}\quad \textrm{(2a),} \quad
        \mathbf{B}_2 = \frac{1}{2}\mathbf{a}_3\quad \textrm{(2a),} \quad
        \mathbf{B}_3 = \frac{1}{3}\mathbf{a}_1 + \frac{2}{3}\mathbf{a}_2 + \frac{1}{4}\mathbf{a}_3 \quad\textrm{(2c),}  \quad \textrm{and} \quad
        \mathbf{B}_4 = \frac{2}{3}\mathbf{a}_1 + \frac{1}{3}\mathbf{a}_2 + \frac{3}{4}\mathbf{a}_3 \quad \textrm{(2c).} \\
\end{equation}
In CrSb (MnTe), Cr (Mn) atoms occupy the 2a sites, while Sb (Te) atoms occupy the 2c sites. The reciprocal-space lattice vectors for the structure are then
\begin{equation}
        \mathbf{b}_1 = \frac{2\pi}{a}\cdot(\frac{2\sqrt{3}}{3},\quad 0,\quad 0), \qquad
        \mathbf{b}_2 = \frac{2\pi}{a} \cdot (\frac{\sqrt{3}}{3},\quad 1,\quad 0),\quad \textrm{and} \qquad
        \mathbf{b}_3 = \frac{2\pi}{c} \cdot (0,\quad 0,\quad 1).
\end{equation}

\section{Electronic band degeneracies in paramagnetic and altermagnetic states}
\label{suppsec:symmetries}

\begin{table}
\caption{Summary of the band structure degeneracy results obtained for altermagnetic and paramagnetic states in the $k_z=0$ and $k_z = \pi/c$ planes as derived in the text.}\label{table:symmetry_degeneracies}
\begin{ruledtabular}
\begin{tabular}{ccc}
{Magnetic state} 
& \({k_z=0}\) plane 
& \({k_z=\pi/c}\) plane \\ \midrule
{Altermagnetic} 
& \(2\times\) degenerate from \(\Theta_{2z} = S_{2z}\,\mathcal T\) 
& \(2\times\) degenerate from \(\widetilde M_z\) \\[4pt]
{Paramagnetic} 
& \(2\times\) degenerate from \(\mathcal{PT}\) 
& \({4\times}\) degenerate from \(\mathcal{PT}\) and \(\widetilde M_z\)\\
\end{tabular}
\end{ruledtabular}
\end{table}

In Figs.~1 and 2 of the main text, and in Figures~\ref{fig:mnte_lsda_electronic_structure} and \ref{fig:mnte_sts_electronic_structure} here, we compare the scalar-relativistic electronic structures of CrSb and MnTe in the altermagnetic ground states and in their paramagnetic states, with the latter described within the disordered local moment (DLM) picture.  Although the site-diagonal spin disorder in the paramagnetic state leads to significant broadening of some electronic bands, it can still clearly be seen that, when going from the altermagnetic to the paramagnetic state, the plane $k_z = \pi/c$ gains an additional two-fold degeneracy compared to the $k_z = 0$ plane. This behaviour can be understood from the different symmetries of the two magnetic states (altermagnetic \textit{versus} paramagnetic).

CrSb and MnTe both crystallise in the NiAs structure, hexagonal space group $P6_3/mmc$ (No.~194), pictured in Fig.~\ref{fig:nias_structure_ibz}. In the altermagnetic phase, the symmetry is described by the corresponding magnetic space group, \textit{i.e.}, the subset of operations derived from $P6_3/mmc$ in which certain space-group elements may be combined with time reversal, $\mathcal{T}$, so as to leave the magnetic structure invariant. By contrast, in the paramagnetic state, due to the equivalence of the two magnetic sublattices, the symmetry reduces to the magnetic grey group $P6_3/mmc1'$, where global time reversal is a symmetry and each space-group operation appears both with and without $\mathcal{T}$. The generators of the non-magnetic space group $P6_3/mmc$ can be chosen to be:
\begin{equation}
P6_3/mmc :
E,\quad \mathcal P,\quad C_{3z}^{+},\quad S_{2z}\equiv\{C_{2z}\,|\,0,0,\tfrac12\},\quad C_{x+y}.
\end{equation}
We then define the $z$-glide mirror, which is also a symmetry of the space group, as
\begin{equation}
\widetilde M_z \equiv \{m_z\,|\,0,0,\tfrac12\}=\mathcal P\,S_{2z}.
\end{equation}
Proceeding, the action of each operation in momentum space is then
\begin{equation}
\begin{aligned}
C_{2z}:&(k_x,k_y,k_z)\mapsto(-k_x,-k_y,k_z),\\
m_z:&(k_x,k_y,k_z)\mapsto(k_x,k_y,-k_z),\\
\mathcal T:&(k_x,k_y,k_z)\mapsto(-k_x,-k_y,-k_z),\\
S_{2z}:&(k_x,k_y,k_z)\mapsto(-k_x,-k_y,k_z),
\end{aligned}
\end{equation}
with
\begin{equation}
(\widetilde M_z)^2 = e^{-ik_z c},\qquad
\Theta_{2z}\equiv S_{2z}\mathcal T:\quad \Theta_{2z}^2 = -\,e^{-ik_z c}.
\end{equation}
We note here that both symmetries $\Theta_{2z}$ and $\widetilde M_z$ map $(k_x,k_y,k_z)\!\to\!(k_x,k_y,-k_z)$, so they leave both planes $k_z=0$ and $k_z=\pi/c$ invariant.

Since $\Theta_{2z}$ is a non-symmorphic symmetry (combined with fractional translation), it acts differently on the planes
$k_z = 0$ or $k_z = \pi/c$,
\begin{equation}
\begin{cases}
k_z=0:\ \ \Theta_{2z}^2=-1 \ \Rightarrow\ \text{two-fold degeneracy across the whole plane},\\[4pt]
k_z=\pi/c:\ \ \Theta_{2z}^2=+1 \ \Rightarrow\ \text{no pairing from }\Theta_{2z}.
\end{cases}
\end{equation}
The same can be said about the glide mirror $\widetilde M_z$ where we obtain that,
\begin{equation}
\begin{cases}
k_z=0:\ \ \widetilde M_z^2=+1 \ \Rightarrow\ \text{no enforced two-fold from }\widetilde M_z\ \text{alone},\\[4pt]
k_z=\pi/c:\ \ \widetilde M_z^2=-1 \ \Rightarrow\ 
\text{two-fold band sticking across the whole plane}.
\end{cases}
\end{equation}
In the altermagnetic state, both of these non-symmorphic symmetries, namely $\Theta_{2z}$ and $\widetilde M_z$, are present. This means that both planes $k_z = 0$ and $k_z = \pi/c$ are two-fold degenerate. In the case of $k_z = 0$, it is $\Theta_{2z}$ that enforces it, whereas for $k_z = \pi/c$ plane, the degeneracy is protected by the glide mirror $\widetilde M_z$. However, in the paramagnetic state all orientations of magnetic moments at each lattice site are equally probable. Thus, global $\mathcal{PT}$ is a symmetry with $(\mathcal{PT})^2=-1$, which implies two-fold degeneracy everywhere (conventional Kramers degeneracy). On $k_z=\pi/c$, the glide-induced sticking due to $\widetilde M_z$ remains which gives an additional independent factor of $2$, leading to an overall four-fold degeneracy on that plane. On $k_z=0$, the $\Theta_{2z}$ symmetry is broken and is no longer a symmetry leading to only two-fold degeneracy along this plane. For convenience, we summarise our degeneracy results regarding the planes $k_z = 0$ and $k_z = \pi/c$ in Table~\ref{table:symmetry_degeneracies}.

\section{Inferring magnetic order: Evaluation of the static, paramagnetic spin susceptibility}

\label{suppsec:spin_susceptibility}

In general, when inferring magnetic ordering in a material, it is desirable to use approaches which are agnostic to the final magnetic ground state. As the DLM picture can describe a paramagnetic state in which there is total magnetic disorder, it is a well-placed starting point for such an agnostic theory.  Since the methodology of the DLM approach to study the onset of magnetic order and to determine magnetic correlations is discussed extensively elsewhere~\cite{gyorffy_first-principles_1985, hughes_onset_2008, mendive-tapia_ab_2019}, with the formal mathematical definitions of key quantities being found in Ref.~\citenum{staunton_static_1986}, we provide only the key details here.

In brief, an estimate of the magnetic two-point correlation function~\cite{frandsen_verification_2016, baral_real-space_2022} and, consequently, the inverse spin susceptibility (Eq.~23 of Ref.~\cite{mendive-tapia_ab_2019}), is given by
\begin{equation}
    \zeta^{\textrm{PM}}_{ij} = 3k_\mathrm{B}T \delta_{ij} + \left. \frac{\partial^2 \bar{\Omega}}{\partial \mathbf{m}_i \partial \mathbf{m}_j} \right|_{\{\mathbf{m}_i\}=\{\mathbf{0}\}} = 3k_\mathrm{B}T \delta_{ij} - S^{(2)}_{ij} = \chi^{-1}_{ij},
    \label{suppeq:susceptibility}
\end{equation}
where $-S^{(2)}_{ij}$ is shorthand for the second partial derivatives of the DFT-DLM internal energy, $\bar{\Omega}$,  discussed in the introduction to this document, with respect to inhomogeneous magnetic order parameter variables $\{\mathbf{m}_i\}$. Owing to the crystal symmetry, it is convenient to work with the lattice Fourier transform of this quantity, $\tilde{\zeta}^{\textrm{PM}}_{ss'}(\mathbf{q})$, defined in Eq.~26 of Ref.~\cite{mendive-tapia_ab_2019}. When searching for a magnetic phase transition, we consider gradually decreasing temperature from the high-$T$ limit and seeking the temperature at which the magnetic susceptibility diverges. Within the present framework, this will be the temperature at which the lowest-lying eigenvalue of $\tilde{\zeta}^{\textrm{PM}}_{ss'}(\mathbf{q})$ passes through zero for some $\mathbf{q}_\textrm{ord}$. The wavevector $\mathbf{q}_\textrm{ord}$ is the \textit{magnetic propagation vector} and defines the nature of ordering between unit cells, while the associated eigenvector defines how the different unit cells in the structure align with one another.

\begin{figure}[t]
    \centering
    \includegraphics[width=0.6\linewidth]{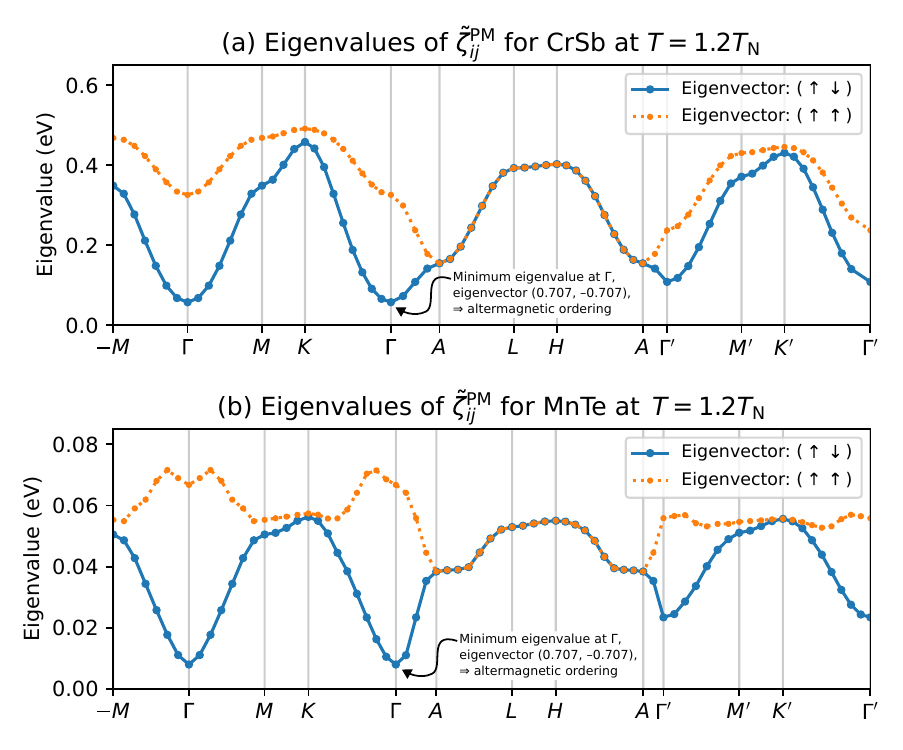}
    \caption{Eigenvalues of $\tilde{\zeta}^{\textrm{PM}}_{ss'}(\mathbf{q})$ at $T=1.2 T_\mathrm{N}$ around high-symmetry lines of the irreducible Brillouin zone for both CrSb (a) and MnTe (b). The quantity $\tilde{\zeta}^{\textrm{PM}}_{ss'}(\mathbf{q})$ is related to an estimate of the static, paramagnetic spin susceptibility of a material. That the minimum occurs at $\Gamma$ with eigenvector $(\uparrow \downarrow)$ indicates that altermagnetic order is predicted for both materials.}
    \label{suppfig:inverse_susceptibility}
\end{figure}

Shown in Fig.~\ref{suppfig:inverse_susceptibility} are plots of the eigenvalues of $\tilde{\zeta}^{\textrm{PM}}_{ss'}(\mathbf{q})$ around the irreducible Brillouin zone of the NiAs structure for both CrSb and MnTe. We note the differing $y$-scales on the two plots; in metallic CrSb, magnetic moments are able to interact strongly with one another via the sea of itinerant electrons, while in semiconducting MnTe, localised magnetic moments interact more weakly. For both CrSb and MnTe, the minimum eigenvalue occurs at the $\Gamma$ point, implying a magnetic propagation vector of $\mathbf{q} = \boldsymbol{0}$ and indicating all unit cells are predicted to have the same sublattice magnetisations. The associated (normalised) eigenvector, defining the sublattice magnetisations is then $(\frac{1}{\sqrt{2}}, -\frac{1}{\sqrt{2}}) = (0.707, -0.707) = (\uparrow \downarrow)$, \textit{i.e.} an altermagnetic ordering. Within this mean-field theory, magnetic ordering is inferred for CrSb and MnTe at temperatures of 1117~K and 160~K, respectively. For CrSb, if an Onsager correction to the mean-field theory~\cite{staunton_onsager_1992} is included, the predicted transition temperature is 920~K, in fair agreement with the experimentally determined value of 705~K~\cite{takei_magnetic_1963}. For MnTe, we attribute the underestimate of $T_\mathrm{N}$ compared to the experimental value of 306~K~\cite{reig_growth_2001} to strong localisation of Mn $3d$ states within the LSIC-LSDA. (Further discussion of this aspect can be found in Sec.~\ref{sec:MnTe_lsda}.)

Additionally, the magnetisation derivatives of Eq.~\eqref{suppeq:susceptibility} can be shown to be related to effective exchange interactions, $J_{ij}$, between magnetic atoms~\cite{mendive-tapia_ab_2019} for the Heisenberg model, which has a Hamiltonian of the form
\begin{equation}
    H(\{\hat{\mathbf{e}}_i\}) = \frac{1}{2}\sum_{i,j} J_{ij} \hat{\mathbf{e}}_i \cdot \hat{\mathbf{e}}_j.
\end{equation} 
 For CrSb, we find that our calculated spin susceptibility data is well-captured by interactions up to fourth-nearest neighbours, with $J_1 = 39.9$~meV, $J_2 = -30.0$~meV, $J_3 = 6.5$~meV,  and $J_4 = 7.2$~meV. For MnTe, a fit up to third-nearest neighbours is sufficient, with $J_1 = 5.0$~meV, $J_2 = -1.8$~meV, and $J_3 = 1.8$~meV. (Here, $J_i$ denotes the exchange interaction between two magnetic atoms which are $i$\textsuperscript{th}-nearest neighbours.) The signs and relative strengths of the dominant interactions compare acceptably with exchange interactions fitted to neutron scattering data for both CrSb~\cite{singh_coherent_nodate} and MnTe~\cite{liu_chiral_2024}, though our calculated values are consistently somewhat larger than those reported experimentally for both materials. We note here, though, that our exchange interactions are recovered from a spin-susceptibility calculation in the paramagnetic regime, and cannot necessarily be expected to be the same as those recovered in the low-temperature altermagnetic regime. (See, \textit{e.g.} Ref.~\citenum{mendive-tapia_short_2021} for a discussion of how DFT-calculated magnetic exchange interactions can vary depending on the choice of reference magnetic state in a calculation.)

\subsection{Degeneracy of spin susceptibilities in the $k_z = \pi/c$ plane}

In Fig.~\ref{suppfig:inverse_susceptibility} above, it is notable that, along the path $A$--$L$--$H$--$A$, \textit{i.e.} in the plane $k_z = \pi/c$, the two eigenvalues of the inverse spin susceptibility matrix are degenerate. These two eigenvalues have corresponding (normalised) eigenvectors $(\frac{1}{\sqrt{2}}, -\frac{1}{\sqrt{2}}) = (0.707, -0.707) =  (\uparrow \downarrow)$ and $(\frac{1}{\sqrt{2}}, \frac{1}{\sqrt{2}}) = (0.707, 0.707) = (\uparrow \uparrow)$. They therefore correspond to the cases where both sublattices spontaneously magnetise in opposite directions (`altermagnetic') and where both sublattices spontaneously magnetise in the same direction (`ferromagnetic'). In the plane $k_z = \pi/c$, we understand the degeneracy of the two eigenvalues as follows.

\begin{figure}[t]
    \centering
    \includegraphics[width=0.3\linewidth]{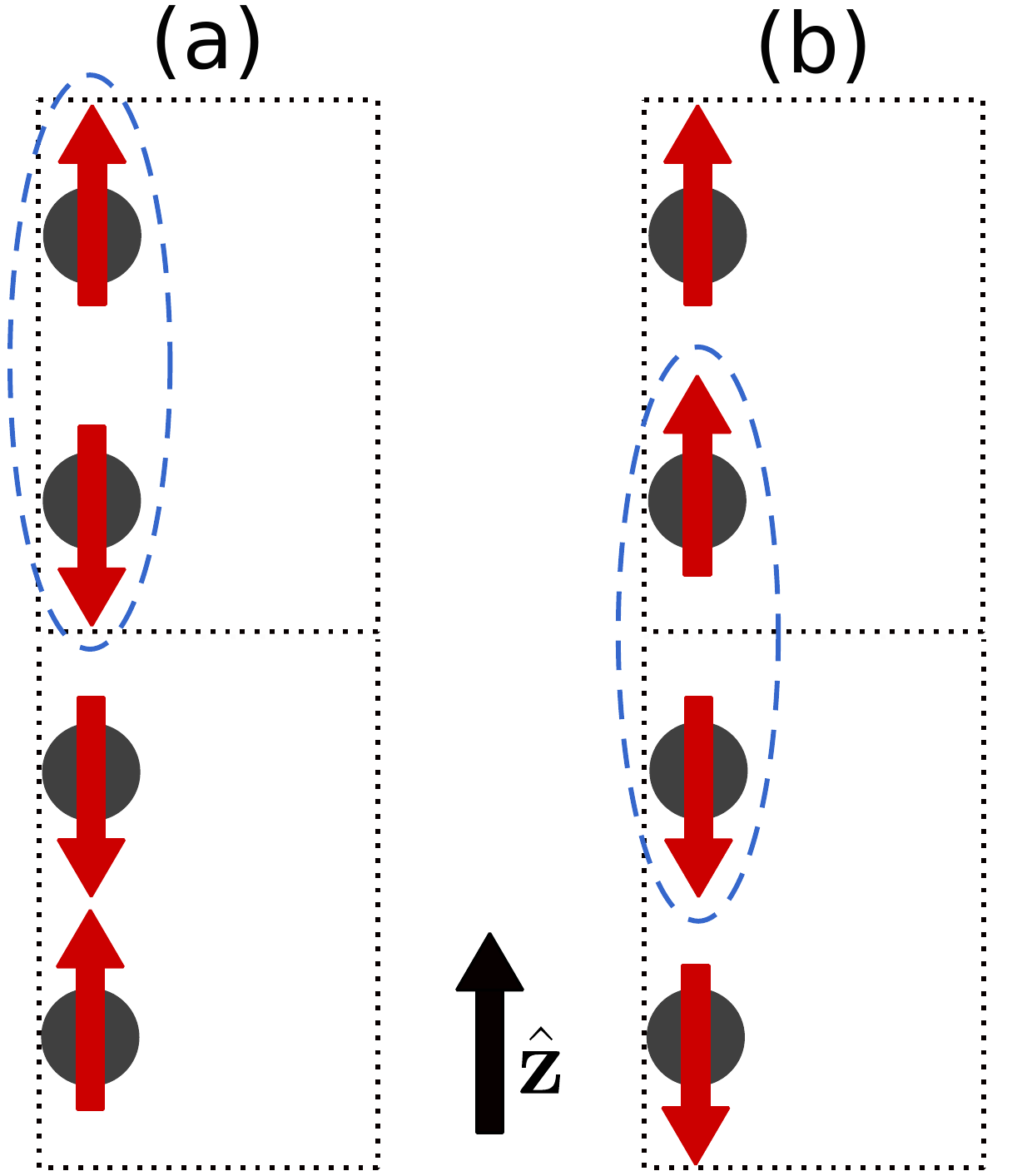}
    \caption{Illustration of how the eigenvectors $(\uparrow,\downarrow)$ and $(\uparrow,\uparrow)$, describing possible antiparallel (a) and parallel (b) spin orderings, respectively, of $\tilde{\zeta}^{\textrm{PM}}_{ss'}(\mathbf{q})$ must have identical eigenvalues when the magnetic propagation vector has $k_z = \pi/c$. Spins from adjacent cells can be `grouped' to produce the same magnetic structure.}
    \label{fig:susceptibility_degeneracy}
\end{figure}

A spin fluctuation with $k_z = \pi/c$ describes a magnetic ordering which is alternating with a periodicity of one unit cell in the $\hat{\mathbf{z}}$ direction. For the NiAs structure considered in this work, `grouping' the atoms between unit cells differently shows that both ferromagnetic and altermagnetic orderings within a single unit cell describe the same overall magnetic structure once the total system is considered. An example of this situation is illustrated conceptually in Fig.~\ref{fig:susceptibility_degeneracy}.

Formally, we consider the non-symmorphic glide mirror
\begin{equation}
\widetilde{M}_z \equiv \{ m_z \,|\, 0,0,\tfrac12 \} = \mathcal{P} \, S_{2z},
\end{equation}
where $m_z$ is a mirror through the $z=0$ plane, $\mathcal{P}$ is spatial inversion, and
$S_{2z} \equiv \{ C_{2z} \,|\, 0,0,\tfrac12 \}$ is a two-fold screw rotation about $z$.
In the conventional, primitive unit cell of CrSb and MnTe, the two magnetic sublattices $A$ and $B$ are exchanged
by $\widetilde{M}_z$.
Let $|A,\mathbf{k}\rangle$ denote the Bloch state on sublattice $A$,
\begin{equation}
|A,\mathbf{k}\rangle = \frac{1}{\sqrt{N}} \sum_{\mathbf{R}}
e^{i\mathbf{k}\cdot(\mathbf{R}+\mathbf{r}_A)} \,
|\mathbf{R}+\mathbf{r}_A\rangle ,
\end{equation}
with $\mathbf{r}_A$ the basis vector of $A$ within the primitive cell.
Applying $\widetilde{M}_z$ yields
\begin{equation}
\widetilde{M}_z \, |A,\mathbf{k}\rangle
= e^{i k_z c/2} \, |B, m_z \mathbf{k}\rangle ,
\end{equation}
and similarly
\begin{equation}
\widetilde{M}_z \, |B,\mathbf{k}\rangle
= e^{i k_z c/2} \, |A, m_z \mathbf{k}\rangle .
\end{equation}
On planes invariant under $m_z$ (e.g., $k_z=0$ and $k_z=\pi/c$),
the Bloch momentum is unchanged ($m_z \mathbf{k} = \mathbf{k}$) and
$\widetilde{M}_z$ has the following $2\times 2$ representation
in the $(A,B)$ sublattice basis:
\begin{equation}
D(\widetilde{M}_z) =
\begin{pmatrix}
0 & e^{i k_z c/2} \\
e^{i k_z c/2} & 0
\end{pmatrix}.
\end{equation}
In particular
\begin{equation}
D(\widetilde{M}_z) =
\begin{cases}
\begin{pmatrix} 0 & 1 \\ 1 & 0 \end{pmatrix} = \sigma_x, & k_z = 0, \\[1.0em]
\begin{pmatrix} 0 & i \\ i & 0 \end{pmatrix} = i\sigma_x, & k_z = \pi/c.
\end{cases}
\end{equation}

To connect this glide to the paramagnetic susceptibility, we introduce the symmetric
($\uparrow \uparrow$, FM-like) and antisymmetric ($\uparrow \downarrow$, AFM-like) sublattice combinations expressed in the $(A,B)$ sublattice basis:
\begin{equation}
v_{\uparrow \uparrow}=\frac{1}{\sqrt{2}}
\begin{pmatrix}
1\\[4pt]1
\end{pmatrix},
\qquad
v_{\uparrow \downarrow}=\frac{1}{\sqrt{2}}
\begin{pmatrix}
1\\[4pt]-1
\end{pmatrix},
\end{equation}
and the unitary change of basis
\begin{equation}
H=\frac{1}{\sqrt{2}}
\begin{pmatrix}
1 & 1 \\[4pt]
1 & -1
\end{pmatrix}.
\end{equation}
In the basis $(v_{\uparrow \uparrow},v_{\uparrow \downarrow})$ the glide is diagonal:
\begin{equation}
D'(\widetilde M_z)\equiv H^\dagger D(\widetilde M_z)\,H =
\begin{cases}
\sigma_z, & k_z=0,\\[4pt]
i\,\sigma_z, & k_z=\pi/c.
\end{cases}
\end{equation}
Hence the explicit action is
\begin{equation}
k_z=0:\quad D'(\widetilde M_z)\,v_{\uparrow \uparrow}=v_{\uparrow \uparrow},\qquad
D'(\widetilde M_z)\,v_{\uparrow \downarrow}=-\,v_{\uparrow \downarrow},
\end{equation}
\begin{equation}
k_z=\pi/c:\quad D'(\widetilde M_z)\,v_{\uparrow \uparrow}=i\,v_{\uparrow \uparrow},\qquad
D'(\widetilde M_z)\,v_{\uparrow \downarrow}=-\,i\,v_{\uparrow \downarrow}.
\end{equation}
The glide invariance
\(
D(\widetilde M_z)\,\tilde\zeta_{ss'}^{\rm PM}(\mathbf q)\,D(\widetilde M_z)^{-1}=\tilde\zeta_{ss'}^{\rm PM}(\mathbf q)
\)
implies that, in the \(\{v_{\uparrow \uparrow},v_{\uparrow \downarrow}\}\) basis, the paramagnetic inverse susceptibility is diagonal for any glide-invariant \(\mathbf q\):
\begin{equation}
\tilde\zeta_{ss'}^{\rm PM}(\mathbf q)=
\begin{pmatrix}
\lambda_{\uparrow \uparrow}(\mathbf q) & 0\\
0 & \lambda_{\uparrow \downarrow}(\mathbf q)
\end{pmatrix}.
\end{equation}

On the plane \(k_z=0\), and since the glide labels are real \((+1,-1)\), symmetry imposes diagonality but does not relate \(\lambda_{\uparrow \uparrow}\) and \(\lambda_{\uparrow \downarrow}\). In general \(\lambda_{\uparrow \uparrow}(\mathbf q)\neq \lambda_{\uparrow \downarrow}(\mathbf q)\).\\

On the $k_z=\pi/c$ plane, the only $\mathbf q$-dependence of $D(\widetilde M_z;\mathbf q)$
is the Bloch phase $e^{i q_z c/2}$ from the half-translation, so
\begin{equation}
D'(\widetilde M_z;-\mathbf q)=D'(\widetilde M_z;\mathbf q)^{*}.
\end{equation}
Thus
\begin{equation}
D'(\widetilde M_z;\mathbf q)=i\sigma_z,\qquad D'(\widetilde M_z;-\mathbf q)=-i\sigma_z.
\end{equation}
Hence
\begin{equation}
D'(\widetilde M_z;\mathbf q)\,v_{\uparrow \uparrow}=i\,v_{\uparrow \uparrow},\quad
D'(\widetilde M_z;\mathbf q)\,v_{\uparrow \downarrow}=-i\,v_{\uparrow \downarrow},
\end{equation}
while
\begin{equation}
D'(\widetilde M_z;-\mathbf q)\,v_{\uparrow \uparrow}=-i\,v_{\uparrow \uparrow},\quad
D'(\widetilde M_z;-\mathbf q)\,v_{\uparrow \downarrow}=i\,v_{\uparrow \downarrow}.
\end{equation}
Thus the $+i$ glide sector is spanned by $v_{\uparrow \uparrow}$ at $\mathbf q$ and by $v_{\uparrow \downarrow}$ at $-\mathbf q$.
Since $\tilde\zeta_{ss'}^{\rm PM}(\mathbf q)=\tilde\zeta_{ss'}^{\rm PM}(-\mathbf q)$ by time-reversal symmetry in the paramagnetic state, we get that $\mathcal T\,\tilde\zeta_{ss'}^{\rm PM}(\mathbf q)\,\mathcal T^{-1}=\tilde\zeta_{ss'}^{\rm PM}(-\mathbf q) = \tilde\zeta_{ss'}^{\rm PM}(\mathbf q)$, and the labels swap under
$\mathbf q\!\to\!-\mathbf q$, one has
\begin{equation}
\lambda_{\uparrow \uparrow}(-\mathbf q)=\lambda_{\uparrow \downarrow}(\mathbf q),\qquad
\lambda_{\uparrow \downarrow}(-\mathbf q)=\lambda_{\uparrow \uparrow}(\mathbf q).
\end{equation}
Together with $\lambda_{\uparrow \uparrow}(\mathbf q)=\lambda_{\uparrow \uparrow}(-\mathbf q)$ (see Fig.~\ref{suppfig:inverse_susceptibility} for evidence of this behaviour along the line $-M$-$\Gamma$-$M$), this gives
\begin{equation}
\lambda_{\uparrow \uparrow}(\mathbf q)=\lambda_{\uparrow \downarrow}(\mathbf q)\qquad (q_z=\pi/c).
\end{equation}
We note that this result follows from the four-fold degeneracy enforced on the $k_z=\pi/c$ plane in the paramagnetic state by the combined $\mathcal{PT}$ symmetry and the nonsymmorphic glide $\widetilde M_z$, as established in Sec.~\ref{suppsec:symmetries}.

\section{Temperature-dependence of the magnetic order parameter}

\label{suppsec:magnetic_order}

In this work, temperature dependence of the magnetic order parameter is described within the DLM picture using the formalism outlined in Ref.~\citenum{staunton_temperature_2006} for describing the temperature dependence of various magnetic quantities \textit{ab initio}. (For description of non-relativistic band splittings, as is the focus of this work, it is sufficient to consider scalar-relativistic calculations and to neglect the calculation of magnetocrystalline anisotropy energies, \textit{etc.}) Concise, accessible summaries of the approach can be found in Sec.~2.3 of Ref.~\citenum{patrick_marmot_2022}, or Sec.~II.B of Ref.~\citenum{mendive-tapia_short_2021}.

As before, the approach considers the statistical physics of thermally fluctuating local magnetic moments. An approximate form of the true grand potential, $\Omega$, is introduced, labelled $\Omega_0$, and defined via
\begin{equation}
    \Omega_0\left(\{\hat{\mathbf{e}}_i\}\right) = -\sum_i \mathbf{h}_i \cdot \hat{\mathbf{e}}_i,
    \label{suppeq:grand_potential}
\end{equation}
where $\mathbf{h}_i$ are the so-called \textit{Weiss fields}, describing the average magnetic field felt locally at a given site $i$. It is then convenient to introduce `beta-h' quantities, $\boldsymbol{\lambda}_i = \beta \mathbf{h}_i$, and also the directional unit vectors, $\hat{\mathbf{n}}_i$ via $\boldsymbol{\lambda}_i = {\lambda}_i \hat{\mathbf{n}}_i$. The form of Eq.~\eqref{suppeq:grand_potential} means that the probability distribution $P_0(\{\hat{\mathbf{e}}_i\})$ can be written as a product of independent probabilities at each lattice site, $P_0(\{\hat{\mathbf{e}}_i\}) = \prod_i P_{0,i}(\hat{\mathbf{e}}_i)$. These independent probabilities at each lattice site can be written analytically as
\begin{equation}
    P_{0,i}(\hat{\mathbf{e}}_i) = \frac{\exp\left( \boldsymbol{\lambda}_i \cdot \hat{\mathbf{e}}_i\right)}{(4\pi /\lambda_i) \sinh(\lambda_i)}.
\end{equation}
The magnetic order parameter at each site, $\mathbf{m}_i = \langle \hat{\mathbf{e}}_i \rangle_{0, T}$, where $0<\| \mathbf{m}_i\| <1$ and $\langle \cdot \rangle_{0, T}$ denotes the ensemble average taken with respect to the free energy of Eq.~\eqref{suppeq:grand_potential}, is then expressed as
\begin{align}
    \mathbf{m}_i &= \int \mathrm{d} \hat{\mathbf{e}}_i \, P_{0,i}(\hat{\mathbf{e}}_i) \hat{\mathbf{e}}_i \, \prod_{j \neq i} \int \mathrm{d} \hat{\mathbf{e}}_j \, P_{0,j}(\hat{\mathbf{e}}_j) \\
    &= L(\lambda_i) \hat{\mathbf{n}}_i,
\end{align}
where $L(x) = \coth{x} - 1/x$ is the usual Langevin function. Naturally, in the case of a two-sublattice, collinear magnetically ordered structure where both sublattices have equal and opposite magnetisation vectors---such as the two altermagnets considered in this work---the magnetic order can be quantified by a single number, $m=\|\mathbf{m}_i\|$.

The temperature dependence of the magnetic order is determined via free energy considerations~\cite{mendive-tapia_short_2021, patrick_marmot_2022}. The Weiss fields are written as
\begin{equation}
    \mathbf{h}_i = -\nabla_{\mathbf{m}_i} \langle \Omega \rangle_{0, T} = -\frac{3}{4\pi} \int \mathrm{d} \hat{\mathbf{e}}_i \langle \Omega \rangle^{\hat{\mathbf{e}}_i}_{0, T} \hat{\mathbf{e}}_i,
    \label{suppeq:weiss_self-consistent}
\end{equation}
where $\langle \cdot \rangle^{\hat{\mathbf{e}}_i}_{0, T}$ denotes a partial average over all possible local moment configurations where $\hat{\mathbf{e}}_i$ is held fixed. Eq.~\eqref{suppeq:weiss_self-consistent} includes an implicit self-consistency condition~\cite{mendive-tapia_short_2021} and, in general, a consistent set $\{\mathbf{h}_i\}$ must be found at each desired temperature to determine the temperature-dependent sublattice magnetisation. (See, \textit{e.g.} Sec.~3.6 of Ref.~\cite{patrick_marmot_2022} for further discussion of this aspect.) In general, Eq.~\eqref{suppeq:weiss_self-consistent} will only admit non-zero solutions below the magnetic critical temperature, the N\'eel temperature in the case of this work.

It is perhaps helpful to think of the above as complementary to the approach outlined in Sec.~\ref{suppsec:spin_susceptibility} for inferring the transition via direct evaluation of the static, paramagnetic spin susceptibility. There, the transition is inferred by considering $T>T_\mathrm{N}$ approaching the N\'eel temperature from `above', while the approach outlined in this section is primarily used (in this work) to consider intermediate temperatures ($0<T<T_\mathrm{N}$) and infer how the well-defined magnetic ground state disorders with increasing temperature, thus approaching $T_\mathrm{N}$ from `below'.

\section{Computational details}

\subsection{The ASA prescription: Insertion of empty spheres and control of sphere radii}

In the Korringa--Kohn--Rostoker (KKR)~\cite{ebert_calculating_2011} formulation of density functional theory (DFT)~\cite{jones_density_2015}, which uses the framework of multiple scattering theory to construct the single-particle Green's function, the one-electron Kohn-Sham effective potential is written as a sum of localised, atom-centred potentials~\cite{faulkner_multiple_2018}. For the calculations presented here, using (where applicable) the local self-interaction correction (LSIC)~\cite{luders_self-interaction_2005} to the local spin-density approximation (LSDA), we use spherically symmetric potentials within the atomic-sphere approximation (ASA) as described by Andersen~\cite{andersen_linear_1975}.

The ASA prescription stipulates that the total volume of the spheres within the unit cell on which the potential is defined is set equal to the volume of the Wigner-Seitz cell of the considered structure. However, for a multi-atom unit cell, such as the NiAs crystal structure considered here, this leaves a level of freedom when it comes to determination of the `optimal' set of sphere radii to use in a given calculation. Additionally, in a comparatively open structure, such as in the NiAs structure, it is commonplace to insert so-called `empty spheres' into the structure where the potential is defined but where there is no associated ionic charge, to improve the space filling and thus the description of the DFT potential~\cite{jepsen_calculated_1995}. The optimal combination of spheres to use will generally be the (physically reasonable) combination which minimises the total energy of the system as obtained within a self-consistent calculation while at the same time avoiding excessive sphere overlap.

For CrSb, we insert two empty spheres into the NiAs structure with fractional coordinates $(\frac{1}{3}, \frac{2}{3}, \frac{3}{4})$ and $(\frac{2}{3}, \frac{1}{3}, \frac{1}{4})$. The sphere radii for spheres associated with Cr, Sb, and `empty' sites are then set to be 1.486, 1.695, and 1.132 \AA, respectively. (So-called `metalloid' elements such as Sb and Te have larger atomic radii than transition metals such as Cr and Mn, and hence larger sphere radii in these calculations.)

For MnTe, with its increased $c/a$ ratio, we find that the best choice is the insertion of four smaller empty spheres with fractional coordinates $(\frac{1}{3}, \frac{2}{3}, \frac{5}{8})$, $(\frac{1}{3}, \frac{2}{3}, \frac{7}{8})$, $(\frac{2}{3}, \frac{1}{3}, \frac{1}{8})$, and $(\frac{2}{3}, \frac{1}{3}, \frac{3}{8})$. The sphere radii for spheres associated with Mn, Te, and `empty' lattice sites are then set to be 1.614, 1.773, and 1.020 \AA, respectively. We find that this setup with four empty spheres results in a lower calculated total energy for MnTe than a setup with two empty spheres, which serves as evidence that this is a more appropriate description of the geometry.

\subsection{Self-consistent DFT calculations}

As was outlined in the main text, we use the all-electron \textsc{Hutsepot} code~\cite{hoffmann_magnetic_2020} to obtain the self-consistent potentials of the Korringa--Kohn--Rostoker (KKR) formulation~\cite{ebert_calculating_2011} of density functional theory (DFT)~\cite{jones_density_2015}. For both materials, experimentally determined lattice constants are used, which for CrSb are  $a=4.121$~\AA\ and $c=5.467$~\AA~\cite{takei_magnetic_1963}, while for MnTe are $a=4.142$~\AA\ and $c=6.711$~\AA~\cite{reig_growth_2001}. We perform scalar-relativistic calculations within the atomic sphere approximation (ASA)~\cite{andersen_linear_1975}, employing an angular momentum cutoff of $l_\text{max} = 3$ for basis set expansions, and a 24-point semi-circular contour in the complex plane to integrate over valence energies. For CrSb and MnTe respectively, we use dense $30 \times 30 \times 24$ and $30 \times 30 \times 20$ $\mathbf{k}$-point meshes for Brillouin zone integrations. 

For CrSb, which is metallic, we use a local spin-density (LSDA) exchange correlation functional, specifically the parametrisation of Perdew and Wang~\cite{perdew_accurate_1992}. For MnTe, which is semiconducting, we employ the local self-interaction correction (LSIC)~\cite{luders_self-interaction_2005} to the LSDA as implemented in \textsc{Hutsepot}~\cite{hoffmann_magnetic_2020} to capture the stronger electronic correlations present in this material, with the LSIC configuration specified below, in Sec.~\ref{sec:lsic_configuration}. For both materials, the altermagnetic ground state is described using a conventional, collinear-spin calculation, while the paramagnetic state is described via application of the DLM picture~\cite{pindor_disordered_1983, staunton_disordered_1984}, which uses the coherent potential approximation (CPA)~\cite{soven_coherent-potential_1967, stocks_complete_1978} to average over all possible spin orientations, including those which are non-collinear. Lloyd's formula~\cite{zeller_elementary_2004} is used for accurate determination of the Fermi energy, $E_\mathrm{F}$.

\subsubsection{LSIC configuration for MnTe}
\label{sec:lsic_configuration}

In MnTe, for a single atom, we apply the LSIC to all $e_g$ and $t_{2g}$ states in a single spin channel, which follows the prescription of Ref.~\cite{dane_self-interaction_2009} and minimises the total energy of the system. In the altermagnetic ground state, these spin channels are selected to be alternate on the two sublattices, which facilitates convergence to the altermagnetic ground state. In the paramagnetic state, we use the equivalence (in the absence of spin-orbit effects) within the CPA of the angular average over spin orientations with that of the CPA condition of a so-called `Ising' alloy of `up' and `down' magnetic moments~\cite{hughes_onset_2008}. The `up' moments have all $e_g$ and $t_{2g}$ states corrected in the majority spin channel, while the `down' moments have all $e_g$ and $t_{2g}$ states corrected in the minority spin channel. In the altermagnetic ground state, the LSIC-LSDA yields a calculated total energy which is lower than that calculated in the LSDA, with the difference being 4.298 eV/f.u. A similar result holds in the DLM calculations describing the paramagnetic state, where the LSIC-LSDA energy is again lower than the LSDA energy, with the difference this time being 4.226 eV/f.u.

\subsection{Density of states and Bloch spectral function calculations}

For calculations of the electronic density of states (DOS) and the Bloch spectral function (BSF), we use a dense $\mathbf{k}$-point mesh of $90\times90\times72$ points ($90\times90\times60$ points) for CrSb (MnTe). All calculations are performed with a consistent small imaginary component of the energy ($\mathrm{Im}E = 2$~mRy) to avoid divergences along the real axis, as is standard practice in the KKR method. These calculations sampled a total of 1601 energies with a spacing between sampled energies of 1~mRy. Where spin polarisation of the BSF is evidenced, \textit{e.g.} in Fig.~1 of the main text, we explicitly plot the difference between the spin-up BSF and the spin-down BSF, \textit{i.e.} $\Delta A_B(\mathbf{k},E) = A_B^\uparrow(\mathbf{k},E) - A_B^\downarrow(\mathbf{k},E)$. Spectral functions are shown on a non-linear scale. Specifically, for the total BSF we plot $\tanh(A_B/A_0)$, while for the spin-resolved data we plot $\tanh(2\Delta A_B/A_0)$, where $A_0=\frac{1}{\pi \mathrm{Im}E}$ is the peak height of a single sharp band.

\subsection{Fermi surface calculations}

The Fermi surface of CrSb is studied by evaluating a $\mathbf{k}$-resolved BSF at a single energy, $E_\mathrm{F}$. We study a constant $k_z$ cut at $k_z=0$ and $k_z = \pi/2c$. The BSF is evaluated for $(k_x, k_y)\in [-0.7, 0.7]^2$, in units of $2\pi/a$. For these calculations, we use $\mathrm{Im}E=0.5$~mRy, to increase the visibility of some of the more subtle features.

\subsection{Evaluation of the static, paramagnetic spin susceptibility}

The static, paramagnetic spin susceptibility, Eq.~\eqref{suppeq:susceptibility}, of both CrSb and MnTe is evaluated within a computational implementation of the framework outlined in Refs.~\cite{gyorffy_first-principles_1985, staunton_static_1986, hughes_onset_2008, mendive-tapia_ab_2019}, with the inclusion of an adaptive meshing scheme for Brillouin zone integrations~\cite{bruno_algorithms_1997}. We sample the derivatives of Eq.~\eqref{suppeq:susceptibility} at a total of 84 points of the irreducible Brillouin zone, including high-symmetry points. From these reciprocal-space data, a least-squares fit is used to recover the effective Heisenberg interactions presented in Sec.~\ref{suppsec:spin_susceptibility}.

\subsection{Temperature dependence of the magnetic order parameter}

The temperature dependence of the magnetic order parameter is described using the formalism of Sec.~\ref{suppsec:magnetic_order} as implemented in the \textsc{Marmot} package~\cite{patrick_marmot_2022}. This code takes self-consistent \textsc{Hutsepot} potentials as inputs and subsequently applies the relevant DLM physics. An adaptive meshing scheme for Brillouin zone integrations is used~\cite{bruno_algorithms_1997}, and angular sampling of magnetic moment orientations uses a $240\times40$ mesh of $\theta$ and $\phi$ values. 

We sample a mesh of input `beta-h' Weiss field values, $\beta \mathbf{h}_i$, (with a denser mesh of points at small beta-h values) and compute their effective temperatures (Eq.~23 of Ref.~\citenum{patrick_marmot_2022}) to obtain the $m(T)$ data shown in Fig.~3 of the main text. As above, here we restate that, by symmetry, the magnetisations of the two sublattices are equal in magnitude and antiparallel for both of our considered materials, so the degree of magnetic order can be quantified by a single number, our altermagnetic order parameter, $m$.

\subsection{The spin asymmetry, $S(m)$}

In the main text, we introduce the spin asymmetry, $S(m)$, in Eq.~1 and Fig.~3(h) to quantify the degree of spin splitting at a given $\mathbf{k}$-point as a function of altermagnetic order parameter, $m$. We evaluate it at $\tilde{\mathbf{k}}$, which is the midpoint of the $\Gamma'$--$M'$ segment of the primed path at fixed $k_z$ on which the altermagnetic splitting is largest. This corresponds to $\tilde{\mathbf{k}} = (0.289, 0, k_z) \frac{2\pi}{a}$, with $k_z = 0.188$ for CrSb and $0.154$ for MnTe. The spin-resolved Bloch spectral functions $A^{\sigma}_{\mathrm{B}}(\tilde{\mathbf{k}},E;m)$ are those of the KKR-CPA calculations described above, evaluated on a uniform mesh of 1601 energies with an imaginary part $\mathrm{Im}\,E = 2$~mRy. These Bloch spectral functions, for all 21 values of $m$, are also shown as the colour maps of Figs.~3(i) and 3(j).

The weight function, $w(E) = \exp[-(E-E_{\mathrm{F}})^2/2\sigma^2]$, with $\sigma = 0.75$~eV effectively restricts the integrals to select states within roughly $\pm 2$~eV of the Fermi level, $E_\mathrm{F}$. The normalisation of $w$ drops out because $S$ is a ratio; $S(0) = 0$ in the paramagnetic state because $A^{\uparrow}_{\mathrm{B}} = A^{\downarrow}_{\mathrm{B}}$ at every $\mathbf{k}$ when $m=0$. In the case of insulating MnTe, where a Fermi level positioned anywhere in the gap would yield the same electron count, the LSIC-LSDA and STS $S(m)$ calculations determine $E_\mathrm{F}$ via measuring the valence-band onset of the density of states at the same $m$. Metallic CrSb and the gapless LSDA description of MnTe have no need for this procedure and take the self-consistently determined $E_{\mathrm{F}}$ values.

The shaded regions in the $S(m)$ panels demonstrate the sensitivity of $S$ to the choice of weight function, thus assessing the robustness of the quantity. We recompute $S$ for nine windows in which the Gaussian width, $\sigma$, is scaled by factors of $0.75$, $1$ and $1.25$ and its centre is shifted by $-0.5$, $0$ and $+0.5$~eV with respect to $E_{\mathrm{F}}$. For the insulating LSIC-LSDA and STS series, where $E_{\mathrm{F}}$ sits at the valence-band maximum, the shifts are chosen to be $-0.5$, $-0.25$ and $0$~eV to keep the window centred on the valence band. The shaded regions on the plots are taken by considering minimum (infimum) and maximum (supremum) of values of $S$ across the nine chosen window functions at each $m$.

To check for potential numerical noise in $S$, we also evaluate this quantity at the end points $\Gamma'$ and $M'$ of the line segment, where the spin splitting vanishes by symmetry (see above, Sec.~\ref{suppsec:symmetries}) and $S$ should be exactly zero. The obtained noise floor is below $2 \times 10^{-3}$ at every $m$ for both materials, at least an order of magnitude below the smallest non-zero $S$ value plotted.

At $m = 1$ the spin asymmetry reaches $S = 0.79$ for CrSb and $0.80$ (LSIC-LSDA), $0.84$ (STS) and $0.63$ (LSDA) for MnTe. The corresponding values at $m = 0.5$ are $0.25$, $0.41$, $0.37$ and $0.20$. For completeness, the tabulated $S(m)$, the underlying spectral functions at $\tilde{\mathbf{k}}$ and the window definitions are provided in the open-access dataset accompanying this work~\cite{dataset}.

\section{M\lowercase{n}T\lowercase{e} within the local spin-density approximation (LSDA)}
\label{sec:MnTe_lsda}

It is well-known that the local spin-density approximation (LSDA) to the exchange-correlation (XC) functional within density functional theory (DFT) underestimates band gaps in many semiconductors and insulators~\cite{perdew_density_2009}. In semiconductors and insulators containing mid- to late-$3d$ elements, this failure is primarily due to the unphysical de-localisation of $3d$ orbitals. Such problems can frequently be rectified via use of hybrid DFT XC functionals~\cite{becke_new_1993}, via application of a Hubbard correction (DFT$+U$)~\cite{himmetoglu_hubbard-corrected_2014}, via application of dynamical mean-field theory (DMFT)~\cite{kotliar_electronic_2006}, or via employment of some form of self-interaction correction (SIC)~\cite{perdew_self-interaction_1981}. 

In this work, we use the local self-interaction correction (LSIC)~\cite{dane_self-interaction_2009} to the LSDA in the context of the KKR formulation of DFT. In the main text, it was demonstrated that these LSIC-LSDA calculations recover the band gap of MnTe, producing results in broad agreement with experimentally determined values. Moreover, it was shown that the band gap is largely unaffected by magnetic disorder within this model. Here, we present results for MnTe modelled within the LSDA without use of any form of self-interaction correction, and demonstrate that the LSDA fails to correctly describe the semiconducting nature of this material. All parameters for these self-consistent LSDA calculations are left unchanged from the LSIC-LSDA calculations on this material.

\begin{figure}[h]
    \centering
    \includegraphics[width=0.45\linewidth]{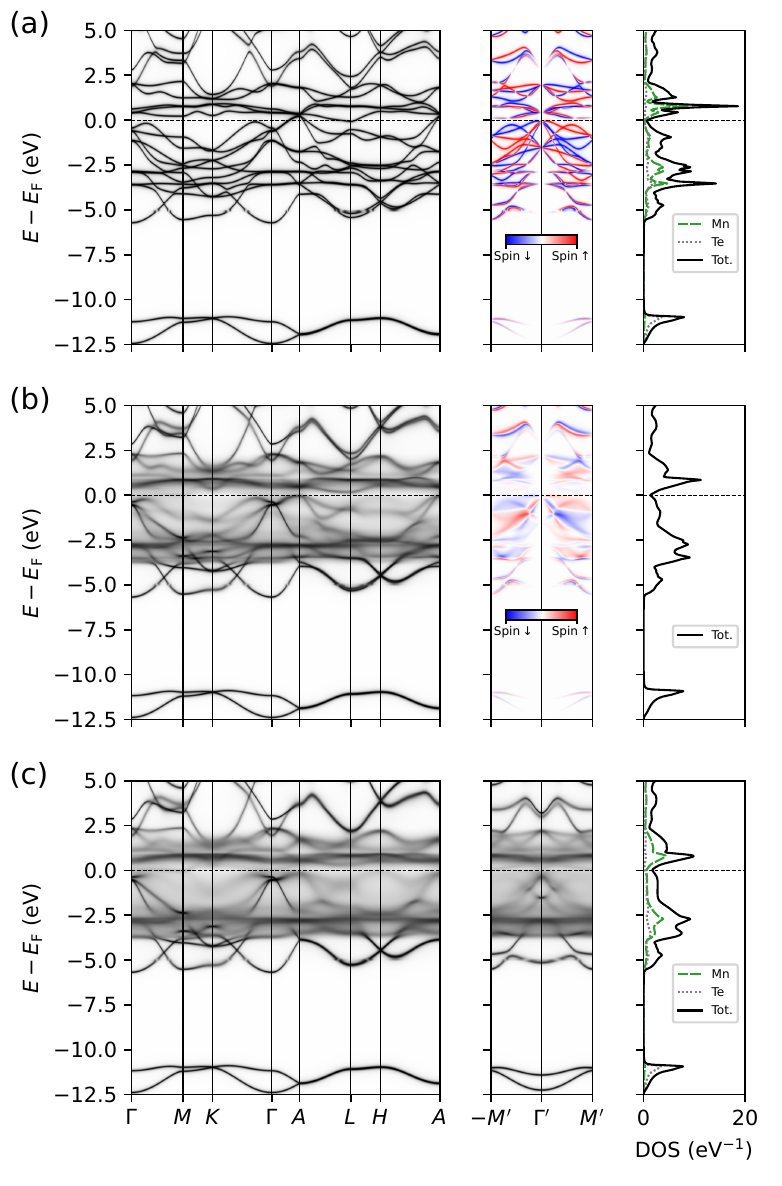}
    \caption{Electronic structure of MnTe within the LSDA in (a) its altermagnetic ground state ($m=1$), (b) at an intermediate value of the altermagnetic order parameter ($m=0.5$ in this case), and (c) its paramagnetic state ($m=0$). Notably, MnTe is metallic in these calculations, at odds with experimental observations and demonstrating the failure of the LSDA to describe the physics of this material.}
    \label{fig:mnte_lsda_electronic_structure}
\end{figure}

\begin{figure}[h]
    \centering
    \includegraphics[width=0.85\linewidth]{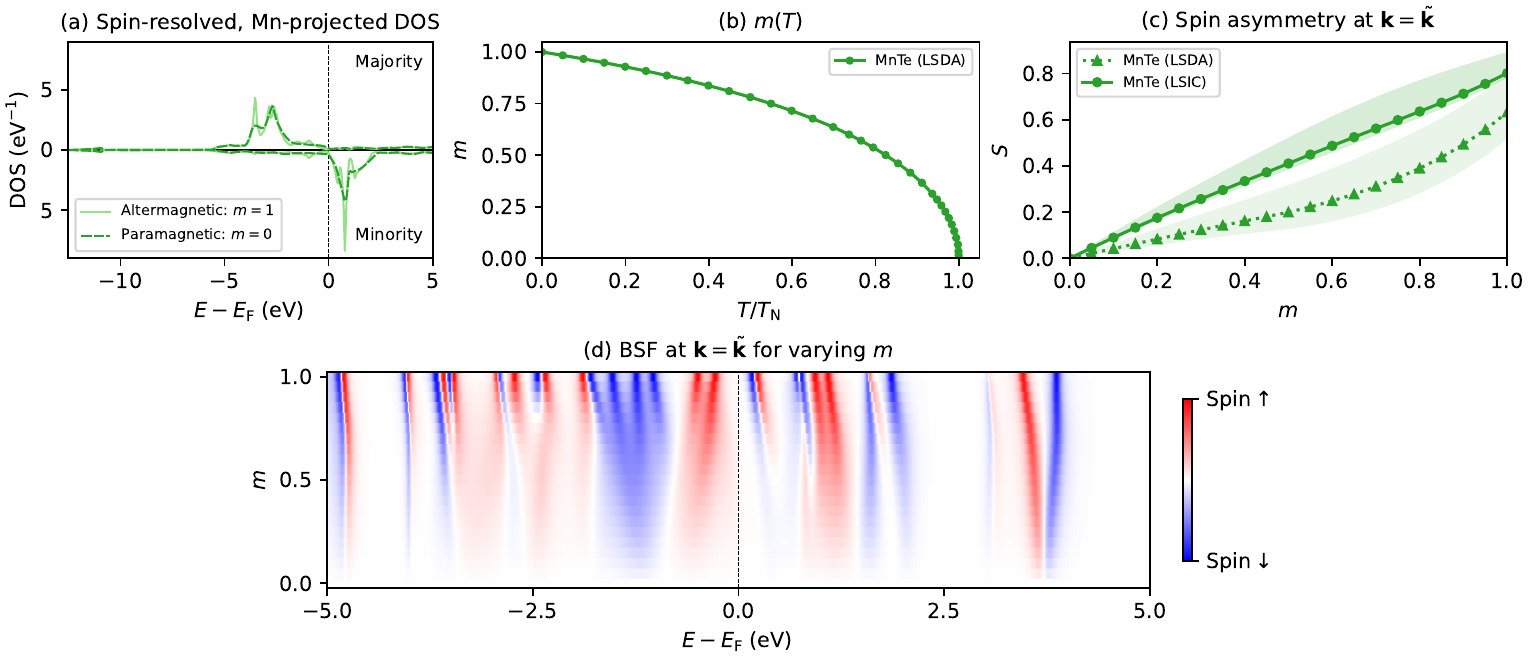}
    \caption{Evolution of the magnetic order and electronic structure of MnTe with increasing temperature as modelled within the LSDA. Panel (a) shows the spin-resolved, species-projected DOS in the limits $m=0$ and $m=1$. Panel (b) shows the altermagnetic order parameter, $m$, as a function of temperature, $T$. Panel (c) shows the evolution of the spin asymmetry, $S$, as a function of altermagnetic order, $m$, together with the LSIC-LSDA result for comparison. Panel (d) shows the evolution of the BSF as a function of $m$ at $\mathbf{k}=\tilde{\mathbf{k}}$.}
    \label{fig:mnte_lsda_evolution}
\end{figure}

First, in Fig.~\ref{fig:mnte_lsda_electronic_structure} we show the electronic structure of MnTe modelled within the LSDA in both altermagnetic (a) and paramagnetic (c) states, as well as for $m=0.5$ (b). In the altermagnetic state (a), the magnitude of the magnetic moment on Mn atoms is now 4.090 $\mu_\mathrm{B}$, while in the paramagnetic (DLM) state (c) this value is 4.122 $\mu_\mathrm{B}$. These values are slightly reduced compared to the LSIC-LSDA values of 4.62 and 4.63 $\mu_\mathrm{B}$ for altermagnetic and paramagnetic (DLM) states, respectively. It can now be seen that, contrary to the LSIC-LSDA results presented in the main text, the $3d$ states of Mn exhibit reduced splitting between spin-up and spin-down states, now occupying an energy range around $E_\mathrm{F}$, resulting in the system being predicted metallic.

Proceeding, we consider the magnetic ordering temperature. Here, the metallic nature of MnTe within the LSDA results in strengthened magnetic interactions. Although the ordering is still predicted to be altermagnetic, with the magnetic propagation vector being $\mathbf{k}=0$ and the eigenvector describing the magnetic ordering being $(\uparrow \downarrow)$, this now occurs at a temperature of 709~K, much higher than the experimental value of 306~K~\cite{reig_growth_2001}. Fitting the real-space interactions for the Heisenberg model to the LSDA spin susceptibility data yields $J_1 = 41.5$~meV, $J_2 = 3.3$~meV, and $J_3 = 11.7$~meV, much larger than values determined from our earlier LSIC-LSDA calculations or from experimental neutron scattering data~\cite{liu_chiral_2024}.

Finally, in Fig.~\ref{fig:mnte_lsda_evolution} we show (a) the contribution to the total electronic density of states made by a single Mn atom within the LSDA in both altermagnetic and paramagnetic (DLM) states, (b) calculated $m(T)$ values for MnTe within the LSDA, with the LSIC results overlaid for comparison, (c) the calculated spin asymmetry, $S(m)$, again with the LSIC results overlaid for comparison, and (d) a slice of the Bloch spectral function around the Fermi level at $\mathbf{k}=\tilde{\mathbf{k}}$ and its evolution as a function of $m$. In this metallic scenario, the spin splitting disappears much more rapidly than in the insulating scenario, suggesting it is the splitting of the $3d$ states to energies above and below $E_\mathrm{F}$ that drives the longer-lasting spin splitting in semiconducting MnTe as compared to metallic CrSb.

\section{M\lowercase{n}T\lowercase{e} within the Slater transition state approximation}
\label{sec:MnTe_sts}

Though the LSIC-LSDA represents a true density functional theory, it should be emphasised that the Kohn--Sham band structure does not necessarily have a ready spectroscopic interpretation. Bands associated with states to which the LSIC is applied are typically pulled down in energy relative to where they might be observed experimentally~\cite{luders_self-interaction_2005}. One route by which to remedy this is via the (non-self-consistent) Slater transition state approximation~\cite{dane_self-interaction_2009}, where only half of the LSIC potential is applied during calculation of quantities such as the Bloch spectral function (BSF) and electronic density of states (DOS). 

\begin{figure}[h]
    \centering
    \includegraphics[width=0.45\linewidth]{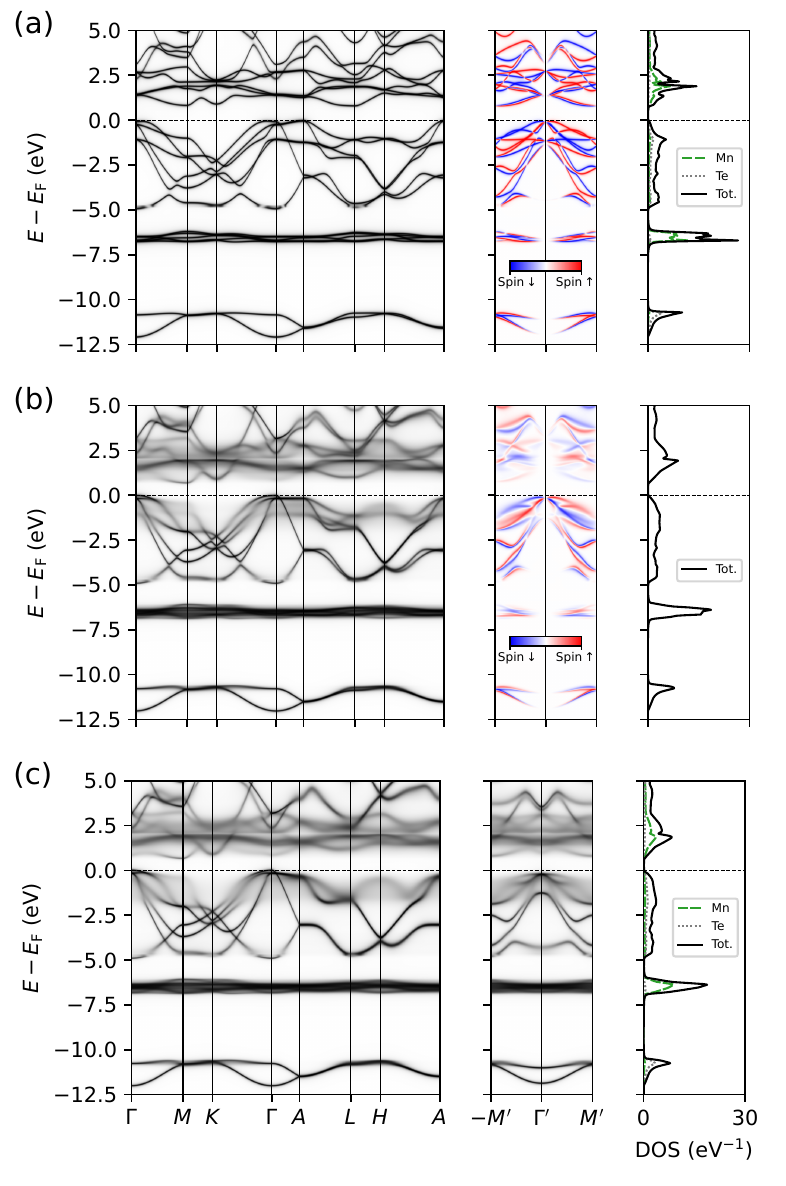}
    \caption{Electronic structure of MnTe within the Slater transition state approximation~\cite{dane_self-interaction_2009} in (a) its altermagnetic ground state ($m=1$), (b) at an intermediate value of the altermagnetic order parameter ($m=0.5$ in this case), and (c) its paramagnetic state ($m=0$). Here, the Mn $3d$ states are split away from the Fermi level, as for the LSIC-LSDA calculations presented in the main text, but not to as low energies.}
    \label{fig:mnte_sts_electronic_structure}
\end{figure}

In Fig.~\ref{fig:mnte_sts_electronic_structure} we show the electronic structure of MnTe modelled within the Slater transition state approximation in both altermagnetic (a) and paramagnetic (c) states, as well as for $m=0.5$ (b). As might be expected, the Mn $3d$ states now sit at energies between those of the LSDA calculation (around the Fermi level) and LSIC-LSDA calculations (between 10 and 12~eV below the Fermi level).

\begin{figure}[h]
    \centering
    \includegraphics[width=0.55\linewidth]{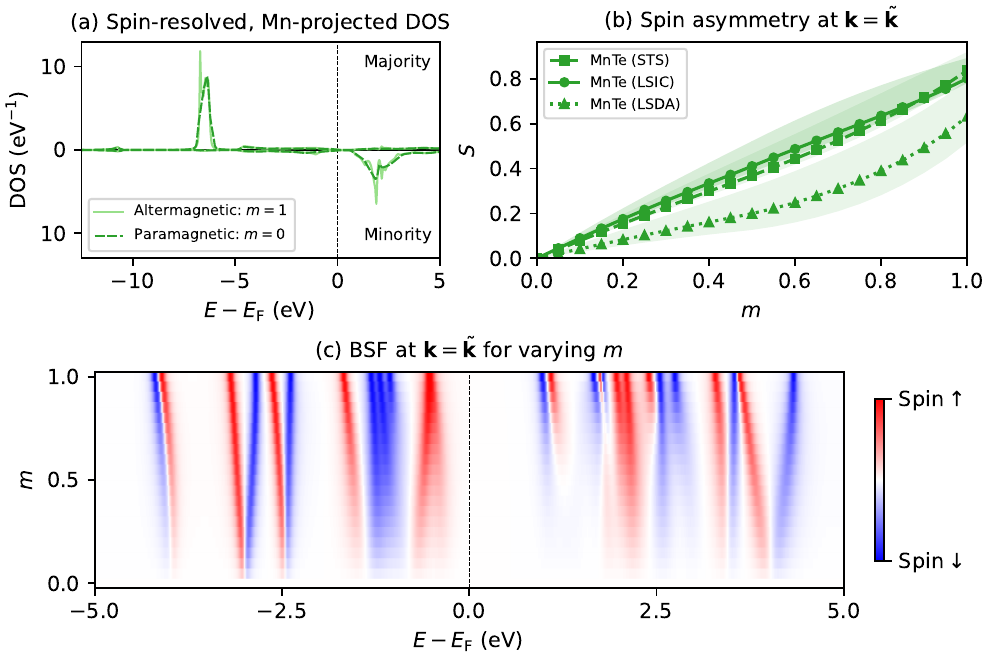}
    \caption{Evolution of the magnetic order and electronic structure of MnTe with increasing temperature as modelled within the Slater transition state approximation. Panel (a) shows the spin-resolved, species-projected DOS in the limits $m=0$ and $m=1$. Panel (b) shows the evolution of the spin asymmetry, $S$, as a function of altermagnetic order, $m$, with LSDA and LSIC-LSDA results for comparison. Panel (c) shows the evolution of the BSF as a function of $m$ at $\mathbf{k}=\tilde{\mathbf{k}}$.}
    \label{fig:mnte_sts_evolution}
\end{figure}

In Fig.~\ref{fig:mnte_sts_evolution} we then show (a) the contribution to the total electronic density of states made by a single Mn atom within the Slater transition state approximation in both altermagnetic and paramagnetic (DLM) states, (b) the calculated spin asymmetry, $S(m)$, with both the LSIC and LSDA results overlaid for comparison, and (c) a slice of the Bloch spectral function around the Fermi level at $\mathbf{k}=\tilde{\mathbf{k}}$ and its evolution as a function of $m$. (Here, $m(T)$ is that of the LSIC-LSDA calculation, since the STS is applied non-self-consistently.) In this scenario where the $3d$ states remain split away from $E_\mathrm{F}$, the spin splitting persists for longer and the $S(m)$ behaviour resembles that of the LSIC-LSDA calculations. This supports our hypothesis that it is the splitting of the $3d$ states to energies above and below $E_\mathrm{F}$ that drives the longer-lasting spin splitting in semiconducting MnTe as compared to metallic CrSb.

\clearpage